\documentclass[a4paper]{jpconf}
\usepackage{graphicx}
\usepackage{amssymb,amsmath,cite}
\usepackage[percent]{overpic}  

\newcommand{\ba}{\begin{eqnarray}}
\newcommand{\ea}{\end{eqnarray}}
\newcommand{\ban}{\begin{eqnarray*}}
\newcommand{\ean}{\end{eqnarray*}}
\newcommand{\bsub}{\begin{subequations}}
\newcommand{\esub}{\end{subequations}}
\newcommand{\R}{\mathbb{R}}

\def\ket#1{|#1\rangle}
	
\def\bsu3{\overline{{\rm SU(3)}}}
\def\bso6{\overline{{\rm SO(6)}}}
\def\bPi2{\overline{\Pi}^{(2)}}

\def\b0{\beta_0}

\def\g0{\gamma_0}

\def\bs{\beta_*}

\begin{document}
\title{Quantum catastrophes from an algebraic perspective}

\author{A Leviatan and N Gavrielov}

\address{Racah Institute of Physics, The Hebrew University, 
Jerusalem 91904, Israel}

\ead{ami@phys.huji.ac.il, noam.gavrielov@mail.huji.ac.il}

\begin{abstract}
  We study the properties of quantum cusp and
  butterfly catastrophes from an algebraic viewpoint.
  The analysis employs an interacting boson model Hamiltonian
  describing quantum phase transitions between specific
  quadrupole shapes by interpolating between
  two incompatible dynamical symmetry limits.
  The classical properties are determined by using coherent
  states to construct the complete phase diagrams associated
  with Landau potentials exhibiting such catastrophes.
  The quantum properties are determined by analyzing the
  spectra, transition rates and symmetry character of the
  eigenstates of critical Hamiltonians.
\end{abstract}
\section{Introduction}

Catastrophe Theory (CT) is a mathematical framework aimed at
exploring the critical points of a family of functions
parameterized by a set of control
parameters~\cite{gilmore-ct,Poston,Stewart82}.
It allows the study
how the critical points (minima, maxima, saddle points),
move about, coalesce and disappear, or bifurcate into new
critical points as the control parameters are varied.
Such an analysis is particularly relevant to the study of
quantum phase transitions (QPTs)~\cite{Gilm78,Gilm79},
which are structural changes in
the properties of a physical system induced by a variation of 
parameters in the quantum Hamiltonian, a topic of great
interest in a variety of
fields~\cite{carr,Sachdev,jolie09,CejJolCas10,iac11}.

Selected quantum properties of elementary catastrophes
(notably, one-dimensional cusp and butterfly)
have been studied by solving a Schr\"odinger equation with
potentials taken from elementary
CT~\cite{gilmore86,Emary05,Cejnar08}.
In the present contribution, we adapt a different approach
and study quantum catastrophes in models based on a
spectrum generating algebra. Such models provide
a rich yet tractable framework for both classical and
quantum treatments of QPTs,
where diverse patterns of structure and symmetries can be
clearly identified. A prototype of such algebraic models is
the interacting boson model (IBM)~\cite{ibm}, describing the
dynamics of quadrupole shapes in nuclei.
Previous CT analysis in this framework, has focused on
classical phase diagrams associated with primarily a cusp
catastrophe~\cite{FengGilmore81,castanos96,ramos03,cejnariac07,
  ramos14,hellemans07} and, to a far less extent, butterfly
catastrophe~\cite{hellemans07,ramos16}.
In what follows, we address both classical and quantum
aspects of cusp and butterfly catastrophes~\cite{gavlev20},
in conjunction with quantum shape-phase transitions
in the IBM framework.

\section{Geometric and algebraic descriptions of quadrupole
  shapes}

A geometric description of quadrupole shapes is based on
an expression of their surface,
\ba
R(\theta,\phi) &=& R_0\, [
  1 + \sum_{m}\alpha_{2,m}Y_{2,m}(\theta,\phi) ] ~,
\label{radius}
\ea
where $R_0$ is the radius of the undistorted spherical surface
and $Y_{2,m}(\theta,\phi)$ are spherical harmonics.
Instead of the $\alpha_{2,m}$, it is convenient to employ an
alternative set of five coordinates involving three Euler
angles $\Omega=(\theta_1,\theta_2,\theta_3)$,
describing the orientation, and two intrinsic variables
$(\beta,\gamma)$ describing the deformation of the shape.
In the body-fixed frame, $\alpha_{2,\pm 1}=0$,
$\alpha_{2,0}=\beta\cos\gamma$ and
$\alpha_{2,\pm 2}=\tfrac{1}{\sqrt{2}}\beta\sin\gamma$.
The radial variable $\beta\geq 0$ measures the total deformation
of the ellipsoid and the angular variable $0\leq\gamma\leq\pi/3$
measures its axiality.
The shape can be spherical $(\beta =0)$ or deformed $(\beta >0)$ with
$\gamma =0$ (axial prolate), $\gamma =\pi/3$ (axial oblate),
$0 < \gamma < \pi/3$ (triaxial), or $\gamma$-independent.
A rotational-invariant Bohr Hamiltonian is constructed from these
generalized coordinates and conjugate momenta, and in its
simplest quantum version has the form~\cite{bohr98},
\ba
\hat{H} = -\frac{\hbar^{2}}{2B} \left [\frac{1}{\beta^4} 
  \frac{\partial}{\partial\beta} \beta^4 \frac{\partial}
       { \partial \beta} 
+\frac{1}{\beta^2\sin 3\gamma}
\frac{\partial}{\partial\gamma}\sin 3\gamma
\frac{\partial}{\partial\gamma}
+\frac{1}{4\beta^2}\sum_{k}\frac{L^2_k}{\sin^2(\gamma
  -\frac{2}{3}\pi k)}\right ] + V(\beta,\gamma) ~.
\label{Bohr}
\ea
The kinetic energy operator is composed of two terms associated
with $\beta$ and $\gamma$ vibrations and a rotational term involving
the angular momentum operators $L_k$.
The potential $V(\beta,\gamma)$ is a function
of two SO(3) invariants: $\beta^2 = \alpha_2\cdot\alpha_2$ and
$\beta^3\cos 3\gamma =
-\sqrt{\tfrac{7}{2}}(\alpha_2\, \alpha_2)^{(2)}\cdot \alpha_2$.
In general, the dynamics of a prescribed quadrupole shape is
revealed by solving a differential equation,
$\hat{H}\Psi(\beta,\gamma,\Omega) =
E \Psi(\beta,\gamma,\Omega)$, which
is the basis of the geometric collective model of
nuclei~\cite{bohr98}.

A particular class of shapes, relevant to the present
contribution, are those arising from Bohr Hamiltonians with
SO(5) symmetry, for which the potential is independent of
$\gamma$, $V(\beta,\gamma)\mapsto V(\beta)$.
In this case, $V(\beta)$ can accommodate either a single minimum
at $\beta=0$ ($\beta>0$), corresponding to a spherical
($\gamma$-unstable deformed) shape or both types of minima
simultaneously, corresponding to coexistence of such shapes.
Representative examples of such $V(\beta)$ are shown in Fig.~1.
\begin{figure}[t]
\centering
\hspace{-0.7cm}
\begin{overpic}[width=5.6cm]{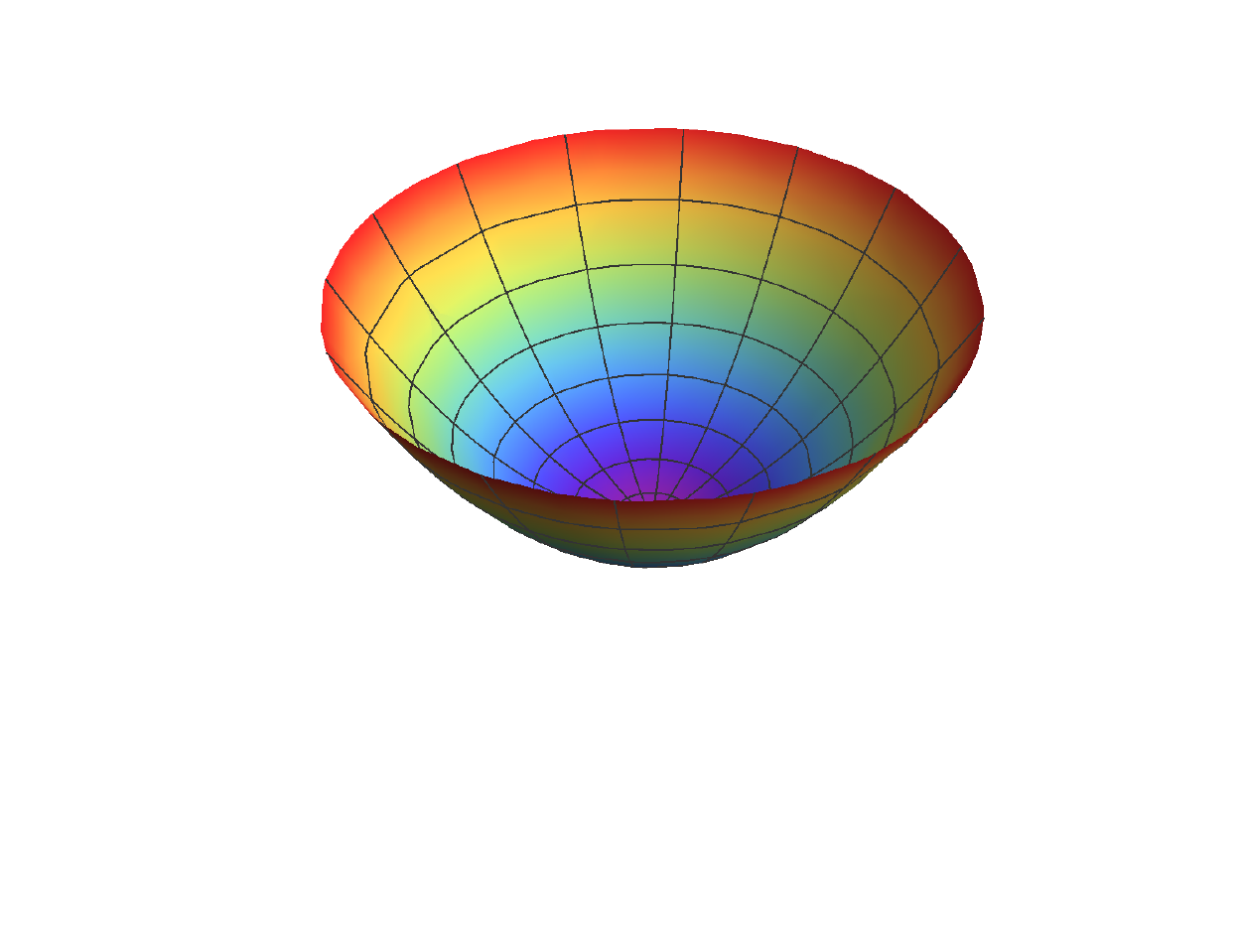}
  \put (46.1,2) {\large(a)}
 \end{overpic} 
  \hspace{-0.4cm}
\begin{overpic}[width=5.6cm]{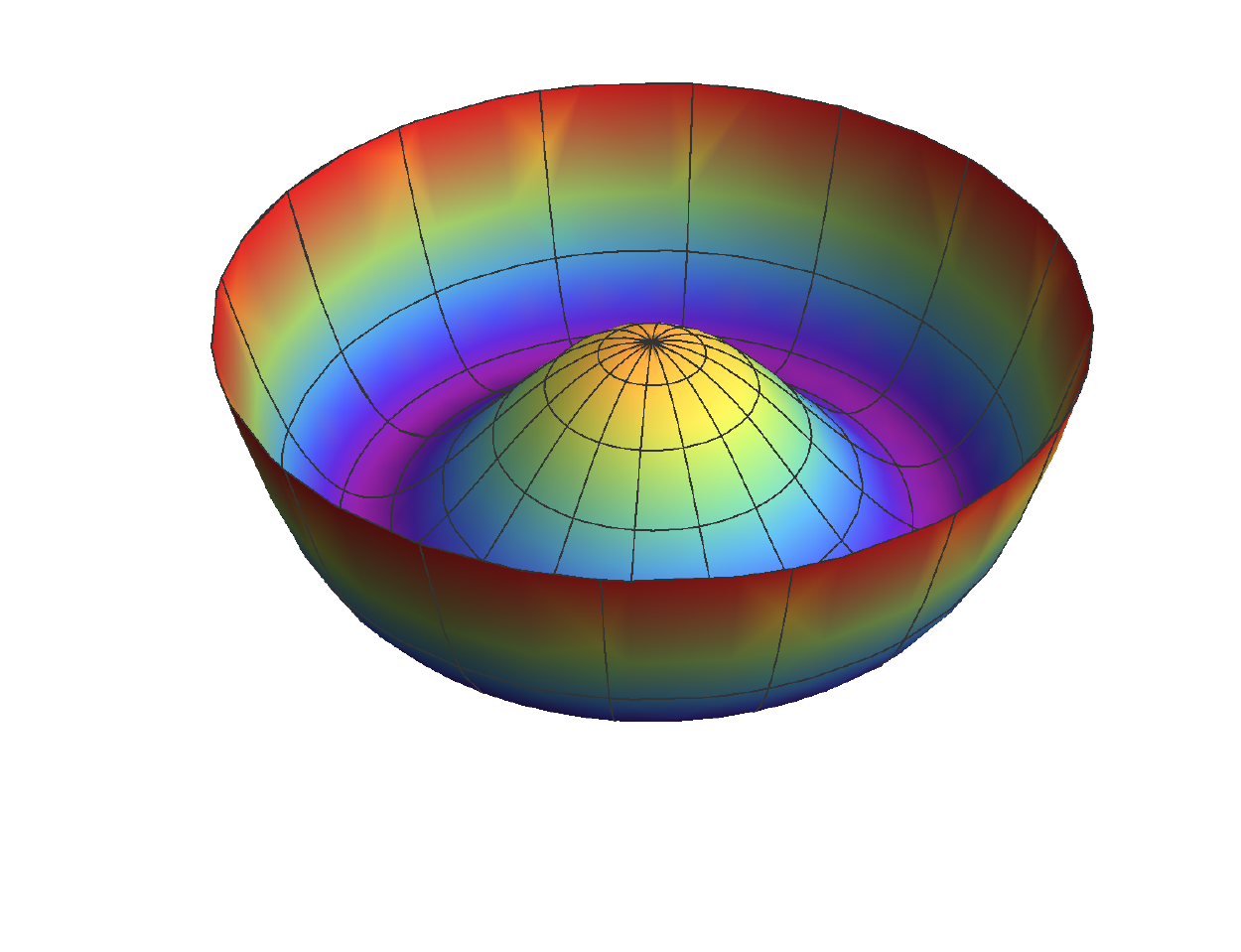}
  \put (48.5,2) {\large(b)}
 \end{overpic} 
  \hspace{-0.4cm}
\begin{overpic}[width=5.6cm]{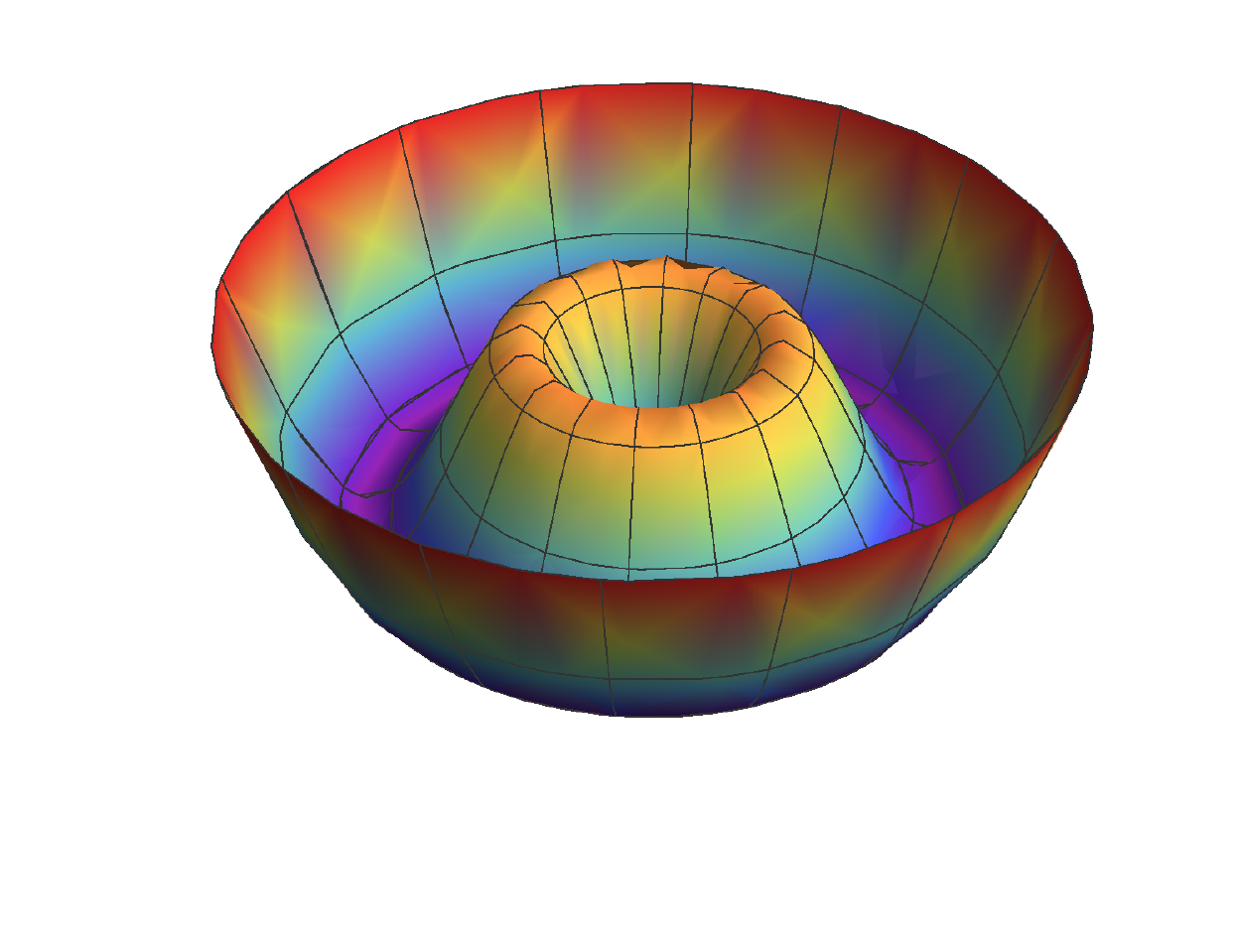}
  \put (50.5,2) {\large(c)}
 \end{overpic} 
  \caption{\label{fig1-3Dpot}
\small
Three-dimensional plots of potentials $V(\beta)$,
in the $(\beta,\gamma)$ variables, accommodating
(a)~a single spherical minimum; (b)~a single deformed minimum;
(c)~coexisting spherical and deformed minima.}
\label{fig1}
\end{figure}

An alternative approach to the dynamics of quadrupole shapes
is that of the interacting boson model (IBM)~\cite{ibm},
describing quadrupole collective states in nuclei in terms of
$N$ interacting monopole $(s)$ and quadrupole $(d)$ bosons.
The model is based on a U(6) spectrum generating algebra with
elements ${\cal G}_{ij}\equiv b^{\dag}_{i}b_j =
\{s^{\dag}s,\,s^{\dag}d_{m},\, d^{\dag}_{m}s,\,d^{\dag}_{m}d_{m '}\}$.
The IBM Hamiltonian is expanded in terms of these generators, 
$\hat{H} = \sum_{ij}\epsilon_{ij}\,{\cal G}_{ij} 
+ \sum_{ijk\ell}u_{ijk\ell}\,{\cal G}_{ij}{\cal G}_{k\ell}$, 
and consists of Hermitian, 
rotational-invariant interactions 
which conserve the total number of $s$- and $d$- bosons, 
$\hat N = \hat{n}_s + \hat{n}_d = 
s^{\dagger}s + \sum_{m}d^{\dagger}_{m}d_{m}$. 
A dynamical symmetry (DS) occurs if the Hamiltonian
can be written in terms of the Casimir operators 
of a chain of nested sub-algebras of U(6).
The Hamiltonian is then completely solvable in the basis
associated with each chain. 
The solvable limits of the IBM correspond to the
following DS chains,
\bsub
\ba
&&{\rm U(6)\supset U(5)\supset SO(5)\supset SO(3)} \;\,\quad
\quad\ket{N,\, n_d,\,\tau,\,n_{\Delta},\,L} \quad 
{\rm spherical\; vibrator}~,\qquad
\label{U5-ds}
\\
&&
{\rm U(6)\supset SO(6)\supset SO(5)\supset SO(3)} \quad
\;\;\;\ket{N,\, \sigma,\,\tau,\,n_{\Delta},\, L} \quad 
\;{\rm \gamma\!-\!unstable\; deformed\; rotor}~,\qquad
\label{SO6-ds}
\\
&&{\rm U(6)\supset SU(3)\supset SO(3)} \;\;\;\qquad\quad\quad
\;\;\,\ket{N,\, (\lambda,\mu),\,K,\, L} \quad
\,{\rm axially\!-\!deformed\;rotor}~.\qquad 
\label{SU3-ds}
\ea
\label{IBMchains}
\esub
The associated analytic solutions 
resemble known limits of the geometric model of
nuclei~\cite{bohr98}, 
as indicated above. The basis members are 
classified by the irreducible representations (irreps)
of the corresponding algebras.
Specifically, the quantum numbers
$N,n_d,(\lambda,\mu),\sigma,\tau,L$,
label the relevant irreps of
U(6), U(5), SU(3), SO(6), SO(5), SO(3), 
respectively, and $n_{\Delta},K$ are multiplicity labels.
Each basis is complete and can be used for a numerical
diagonalization of the Hamiltonian in the general case.

Of particular relevance to the present contribution are
the DS-chains with leading sub-algebras U(5) and SO(6).
Typical spectra and $E2$ rates are shown in Fig.~2.
The U(5)-DS limit of Eq.~(\ref{U5-ds}) 
is appropriate to the dynamics of a spherical shape.
The spectrum resembles that of a spherical vibrator, where
states are arranged in U(5) $n_d$-multiplets with strong
connecting ($n_d+1\!\to\!n_d$) $E2$ transitions.
The lowest U(5) multiplets involve
the ground state with quantum numbers
$(n_d\!=\!0,\,\tau\!=\!0,\, L\!=\!0)$ 
and excited states with quantum numbers 
$(n_d=\!1\!,\,\tau\!=\!1,\, L\!=\!2)$, 
$(n_d\!=\!2:\,\tau\!=\!0,\,L\!=\!0;\,\tau\!=\!2,\,L\!=\!2,4)$ and 
$(n_d\!=\!3:\,\tau\!=\!3,\,L\!=\!0,3,4,6;\,\tau\!=\!1,\,L\!=\!2)$.
The SO(6)-DS limit of Eq.~(\ref{SO6-ds}) is appropriate to the 
dynamics of a $\gamma$-unstable deformed shape. The spectrum
resembles that of a $\gamma$-unstable deformed 
roto-vibrator, where states are arranged 
in SO(6) $\sigma$-multiplets forming rotational bands
with strong connecting ($\tau+1\!\to\! \tau$)
$E2$ transitions between $(\tau,L)$ states in each band.
The lowest irrep $\sigma\!=\!N$ contains the ground ($g$) band
and the first excited irrep $\sigma\!=\!N-2$ contains the
$\beta$-band.
The lowest members in each band have quantum numbers 
$(\tau\!=\!0,\, L\!=\!0)$, $(\tau\!=\!1,\, L\!=\!2)$, 
$(\tau\!=\!2,\, L\!=\!2,4)$ and $(\tau\!=\!3,\, L\!=\!0,3,4,6)$.
\begin{figure}[t!]
\centering
\hspace{-0.7cm}
\begin{overpic}[width=8cm]{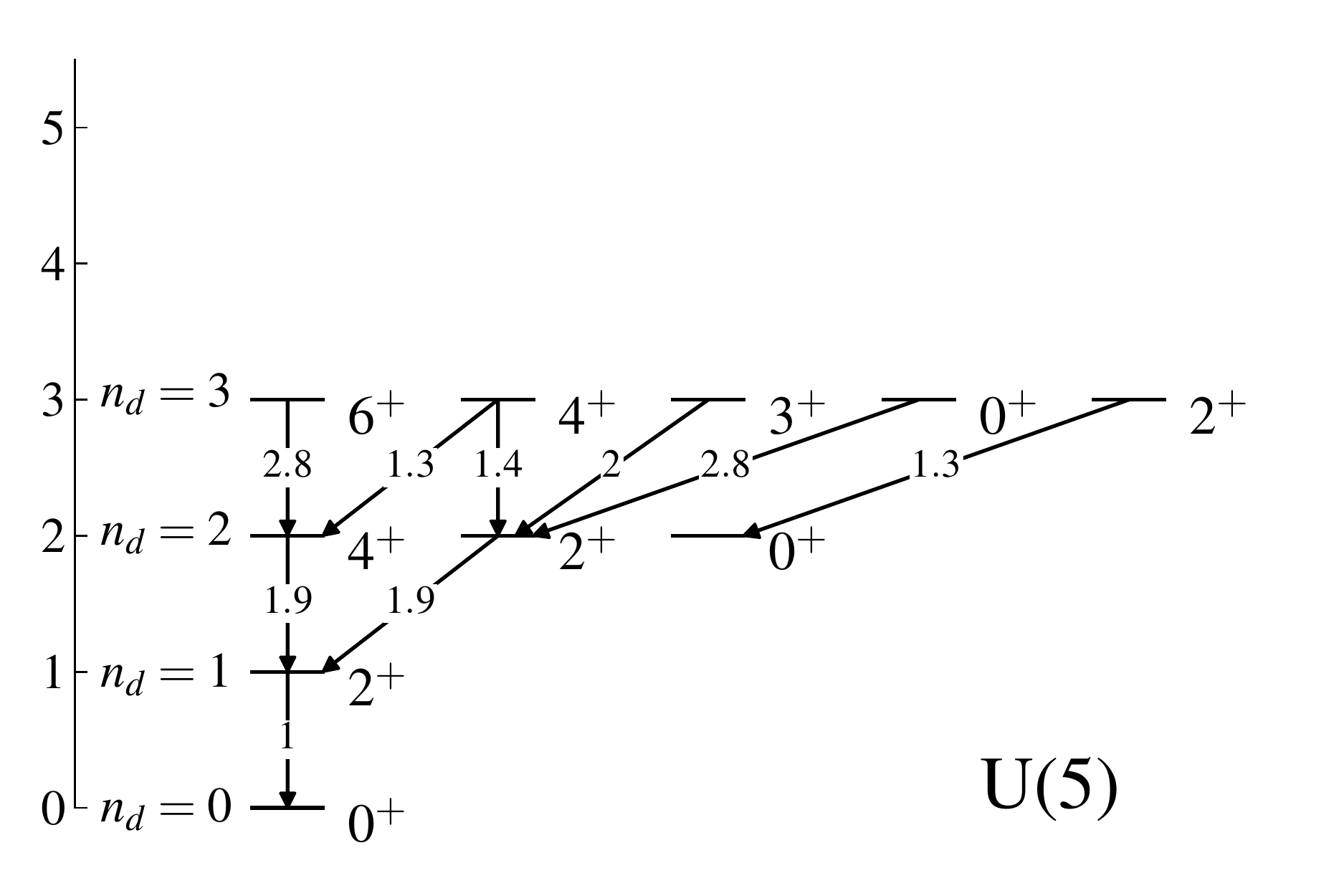}
\put (63,6.5) {\large(a)}
\end{overpic}
\hspace{-0.4cm}
\begin{overpic}[width=8cm]{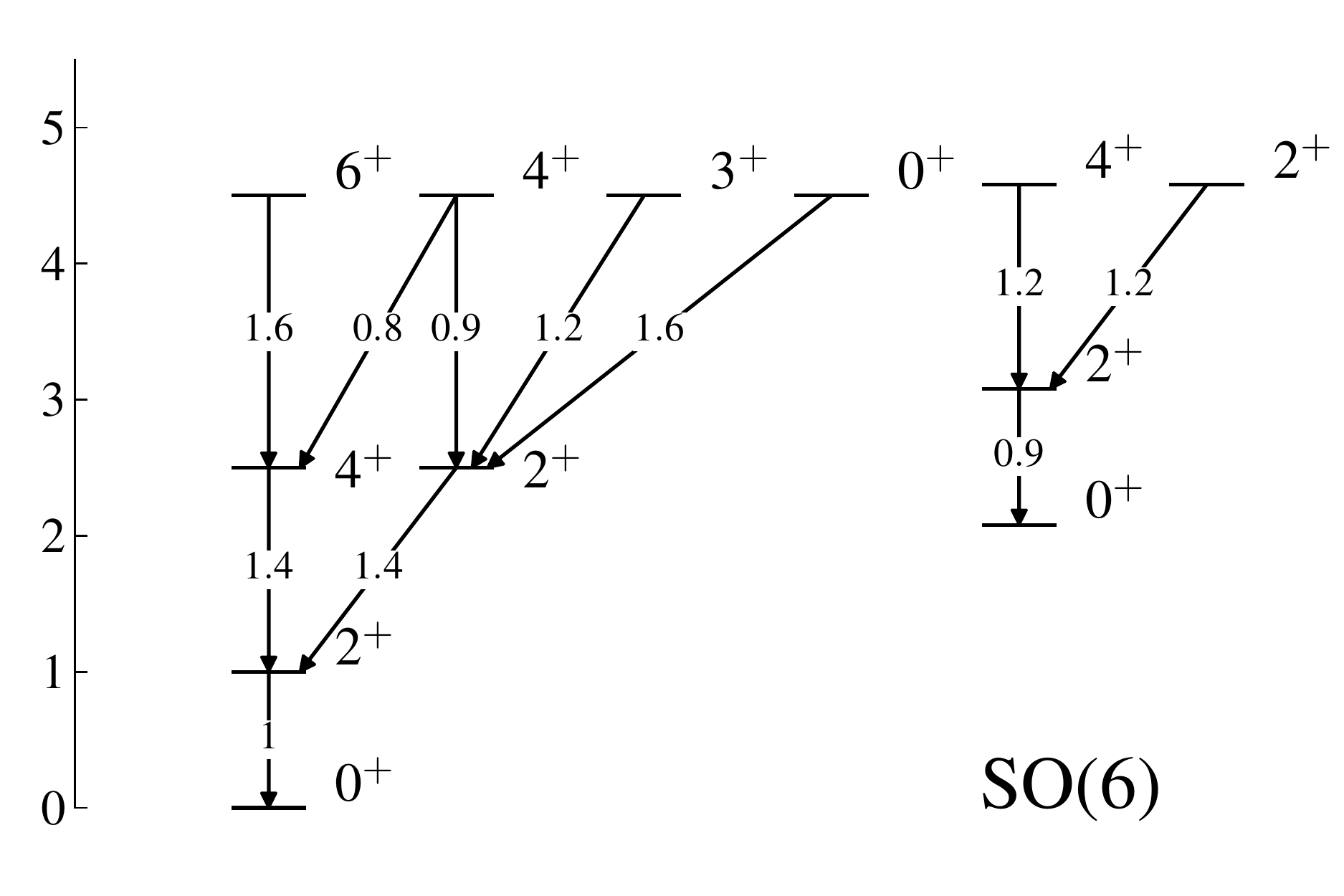}
\put (63,6.5) {\large(b)}
\put (17.5,1) {$\sigma=N$}
\put (72.5,22.5) {$\sigma=N-2$}
\end{overpic}
\caption{\label{fig2-u5so6ds}
\small
Energy spectra [in units of $E(2^{+}_1)=1$] and $E2$ rates
[in units of $B(E2;2^{+}_1\to 0^{+}_1)=1$].
(a)~U(5)-DS limit, Eq.~(\ref{U5-ds}).
(b)~SO(6)-DS limit, Eq.~(\ref{SO6-ds}).
The $E2$ operator is proportional to
$d^{\dag}s + s^{\dag}\tilde{d}$ and $N=25$.}
\label{fig2}
\end{figure}

A geometric visualization of the model is obtained by
means of a coset space 
$U(6)/U(5)\otimes U(1)$  and a `projective' coherent 
state~\cite{gino80,diep80},
\bsub
\ba
\vert\beta,\gamma ; N \rangle &=&
(N!)^{-1/2}(b^{\dagger}_{c})^N\,\vert 0\,\rangle ~,\\
b^{\dagger}_{c} &=& (1+\beta^2)^{-1/2}[\beta\cos\gamma 
d^{\dagger}_{0} + \beta\sin{\gamma} 
( d^{\dagger}_{2} + d^{\dagger}_{-2})/\sqrt{2} + s^{\dagger}] ~,
\ea
\label{int-state}
\esub
from which an energy surface is derived by the 
expectation value of the Hamiltonian,
\ba
E_{N}(\beta,\gamma) &=& 
\langle \beta,\gamma; N\vert \hat{H} \vert \beta,\gamma ; N\rangle ~. 
\label{enesurf}
\ea 
Here $(\beta,\gamma)$ are
quadrupole shape parameters, analogous to those of the
geometric model, Eq.~(\ref{Bohr}). Their values, $(\b0,\g0)$,
at the global minimum of $E_{N}(\beta,\gamma)$ define the
equilibrium shape for a given IBM Hamiltonian. When the
Hamiltonian has SO(5) symmetry, the energy surface is
independent of $\gamma$,
$E_{N}(\beta,\gamma)\mapsto E_{N}(\beta)$.
The equilibrium deformations associated with the 
DS limits, Eq.~(\ref{IBMchains})
conform with their geometric interpretation 
and are given by $\b0\!=\!0$ for U(5), 
$(\b0\!=\!1,\g0\,\,{\rm arbitrary})$ for SO(6) 
and $(\b0 \!=\!\sqrt{2},\g0\!=\!0)$ for SU(3).

\section{Quantum phase transitions}

Quantum phase transitions (QPTs) are qualitative changes in the 
ground state properties of a physical system induced by
a variation of parameters $\xi$, called the control parameters,
in the quantum Hamiltonian $\hat{H}(\xi)$~\cite{Gilm78,Gilm79}.
The particular type of QPT is reflected in the topology of the 
underlying mean-field (Landau) potential $V(\xi;\beta)$.
In a second-order (continuous) QPT, $V(\xi;\beta)$ has
a single minimum $(\xi<\xi_c)$ which evolves continuously as
a function of $\xi$ into another minimum
$(\xi>\xi_c)$. Along the way, the potential becomes flat at the
critical-point $(\xi=\xi_c)$.
In a first-order (discontinuous) QPT, the potential starts
with a single minimum $(\xi<\xi^{*})$ and at the spinodal point
$(\xi=\xi^{*})$ a second local minimum develops. The two
minima cross and become degenerate at the critical-point
$(\xi=\xi_c)$.
For $\xi>\xi_c$, the first minimum becomes local and disappears
at the anti-spinodal point ($\xi=\xi^{**})$. For $\xi>\xi^{**}$,
$V(\xi;\beta)$ has again a single minimum. The range of values,
$\xi^{*}< \xi < \xi^{**}$, defines the coexistence region,
where the potential accommodates simultaneously the two minima.
The above behaviour of the Landau potentials for first- and
second order QPTs, are shown schematically in Fig.~3.
\begin{figure}[t]
\centering
\begin{overpic}[width=16cm]{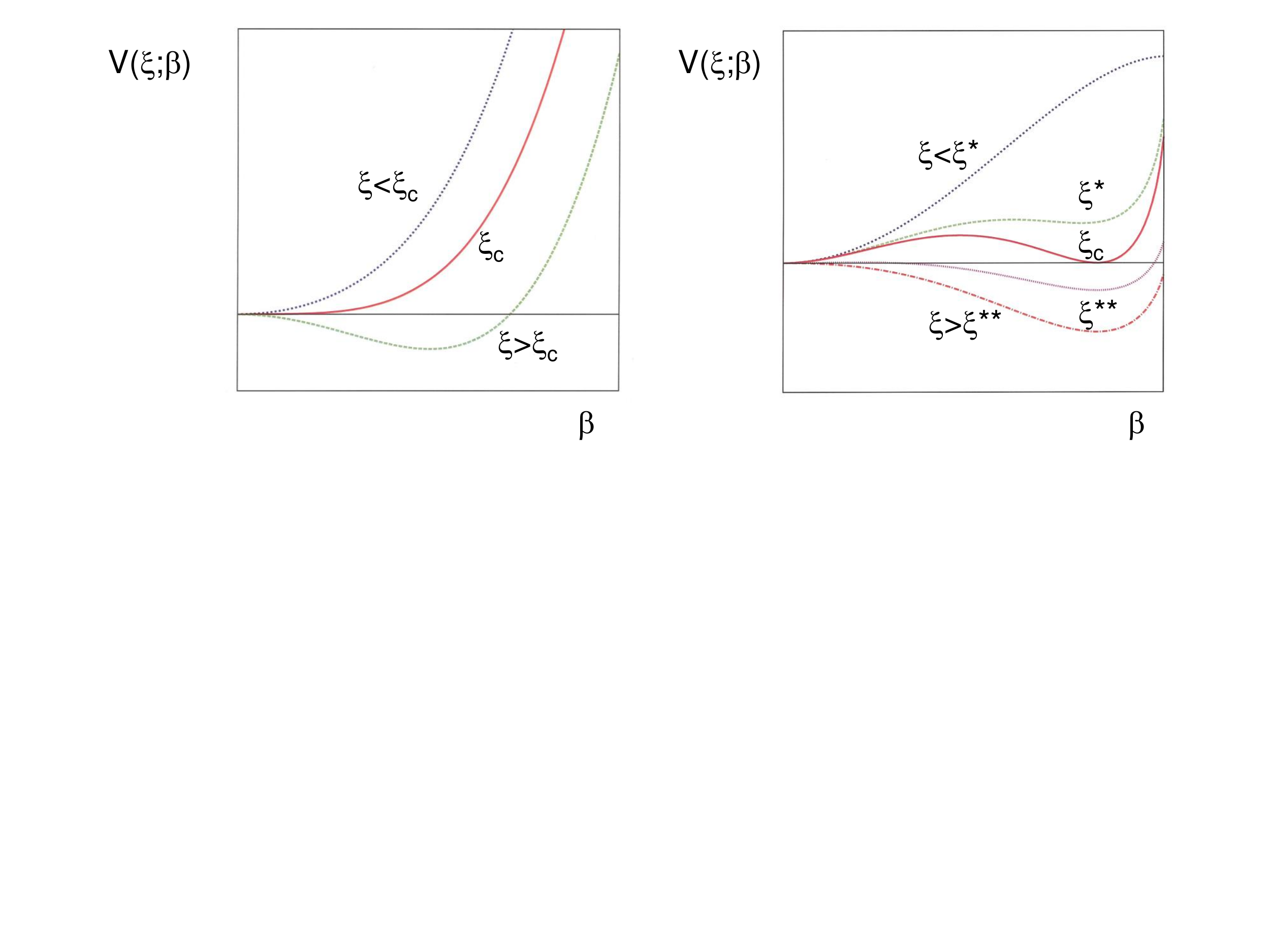}
\put (20,69) {\large(a)}
\put (63,69) {\large(b)}
\end{overpic} 
\vspace{-7cm}
\caption{\label{fig3-Landau-qpt12}
\small
Landau potentials $V(\xi;\beta$) for (a)~second-order and
(b)~first-order transitions. The potentials are labeled by
the value of the control parameter $\xi$. The values
$\xi^{*}$, $\xi_c$ and $\xi^{**}$ are the spinodal, critical
and anti-spinodal values, respectively.}  
\end{figure}

In the algebraic approach to QPTs, the dynamical symmetries
correspond to possible shape-phases of the system
in accord with their geometric interpretation.
The relevant Hamiltonian
mixes terms from different DS
chains~\cite{diep80,iaczamcas98,iaczam04},
\ba
\hat{H}(\xi) = \xi\,\hat{H}_{G_1} + (1-\xi)\,\hat{H}_{G_2} ~,
\label{Hxi}
\ea
where
$G_i\!=\!{\rm U(5),\, SO(6),\,SU(3)}$ and
$G_1\!\neq\! G_2$.
The Hamiltonian interpolates between the different DS limits
of Eq.~(\ref{IBMchains}) by varying the control parameter $\xi$.
The energy surface, $E_{N}(\xi;\beta,\gamma)$ of
Eq.~(\ref{enesurf}), depends parametrically on $\xi$ and
serves as the classical mean-field Landau potential.
The equilibrium deformations $[\b0(\xi),\g0(\xi)]$, where the
surface has a global minimum, serve as the order parameters.
The type of phase transition and critical points
are determined by the derivatives with respect to $\xi$ of
the energy surface, evaluated at the equilibrium deformations.
The order of the QPT is the order of the derivative
where discontinuities first occur (Ehrenfest classification).
IBM Hamiltonians of the above form have been used extensively
for studying shape-phase transitions in
nuclei~\cite{jolie09,iac11,CejJolCas10}.
The terms $\hat{H}_{G_i}$ in Eq.~(\ref{Hxi}) are usually
taken to be the Casimir operators of the leading sub-algebras
in each of the two DS-chains.

As is evident from Fig.~3, the nature of the QPT depends on the
surface landscape (minima, maxima, saddle points) whose
properties change with $\xi$. It is therefore necessary
to perform a detailed analysis of such changes and reveal the
critical behaviour of a~family of functions
$E_{N}(\xi;\beta,\gamma)$ which, in general, depend on several
control parameters ($\xi$) and state variables $(\beta,\gamma)$.
Catastrophe Theory is a mathematical formalism
particularly suitable for such a task.
\begin{figure}[b]
\centering
\includegraphics[width=16cm]{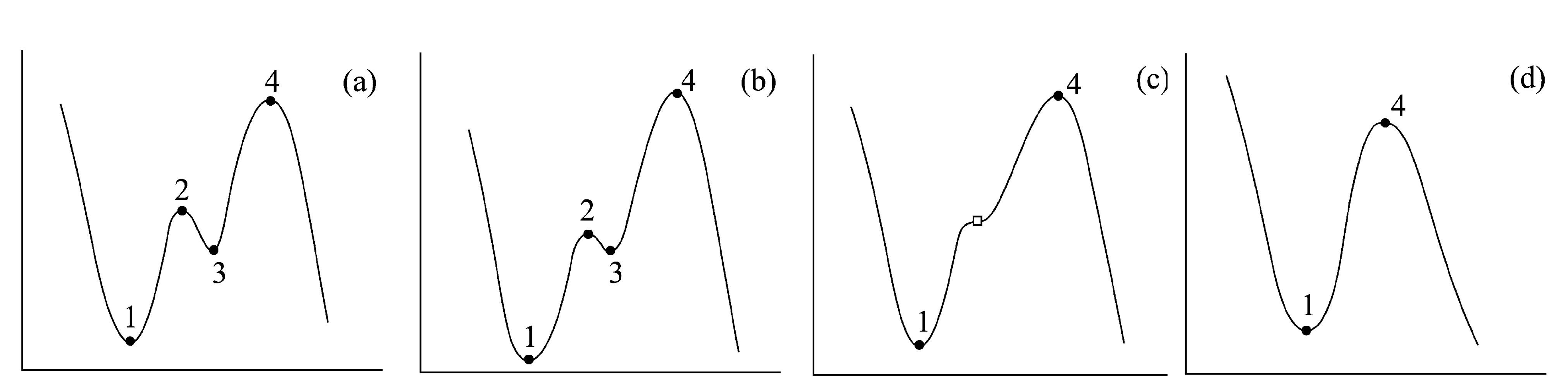}
\caption{\label{fig4-CT}
\small
Illustration of the fact that degenerate critical points
organize the qualitative behaviour of a family of functions.
Panels (a), (b), (c) and (d) show an arbitrary function for
different sets of the control parameters. In panels (a), (b)
and (d) the function exhibits only isolated critical points
(indicated by~\scalebox{0.7}{$\bullet$})
while in panel (c) it has a degenerate critical point
(indicated by \scalebox{0.6}{$\square$}).
The number of isolated critical points or extrema changes
when the control parameters pass through (c) corresponding
to a degenerate critical point.
Adapted from ~\cite{hellemans07}.}
\end{figure}

\section{Elements of Catastrophe Theory}

Catastrophe Theory (CT)~\cite{gilmore-ct,Poston,Stewart82}
provides an efficient tool to
investigate the critical behavior of a family of functions,
$V(x,a)\in {\R}$, which depend on state variables
$x\equiv (x_1,x_2,\ldots,x_n)\in {\R}^n$
and control parameters $a\equiv (a_1,a_2,\ldots,a_k)\in {\R}^k$.
The functions support three types of points.
(i)~Regular points, which have
a nonzero gradient ($\nabla V\neq 0$) and points
where the gradient vanishes, called stationary or critical points,
which fall into two categories:
(ii)~Morse points (isolated non-degenerate critical points),
for which $\nabla V= 0$ and ${\rm det}(V_{ij})\ne 0$
and (iii)~non-Morse points (non-isolated degenerate critical
points), for which $\nabla V= 0$ and ${\rm det}(V_{ij})=0$.
Here ${\rm det}(V_{ij})$ stands for the determinant of
the Hessian (the matrix of second derivatives
$V_{ij} = \frac{\partial^2 V}{\partial x_i\partial x_j}$).

For a particular function, most points $x\in {\R}^n$
are regular and its qualitative behavior is determined by the
isolated non-degenerate critical points whose nature (minima,
maxima, saddle) is dictated by the eigenvalues of the Hessian.
For a family of functions $V(x,a)$
most points $a\in {\R}^k$ parameterize Morse functions,
yet it is the non-Morse functions with degenerate
critical-points, which completely organize the global
properties of the family. This follows from the fact that
stationary points are either created or destroyed at a
non-Morse point.
Consequently, the topology of the function changes drastically
when the control parameters are varied and pass through values
appropriate to a non-Morse function, as illustrated in Fig.~4.

The set in the parameter space ${\R }^k$ that defines
the non-Morse points is called the bifurcation set and
it separates the state space into regions where the
functions have qualitative different behavior.
If two such lines intersect, the degeneracy of the crossing point,
called the triple point, is higher than the degeneracy of the
critical points determining the two lines.
Another set of parameters in ${\R }^k$ ,
called the Maxwell set, defines the ensemble of critical points
with equal values of the function ({\it e.g.}, degenerate minima).
The union of the bifurcation and Maxwell sets is called the
separatrix. These sets partition  ${\R }^k$
into different regions, where the qualitative properties
of the functions $V(x,a)$ remain unchanged.
In the context of phase transitions,
where $V(x,a)$ is the Landau potential, the separatrix
mark out specific boundaries in the phase diagram whose
crossing, upon variation of control parameters,
correspond to a phase transition.

Thom's splitting lemma shows that near a degenerate critical
point, the function $V(x,a)$ can be written as sum of
a Morse part and a non-Morse
part~\cite{gilmore-ct,Poston,Stewart82}. The Morse part can be
brought by a smooth change of variables to a quadratic form.
The non-Morse part can be represented by
a canonical form called a catastrophe function.
The latter function is the sum of a catastrophe germ
containing the non-Morse point and a universal perturbation.
Thom listed all catastrophe functions for $n=1,2$ state variables
and $k\leq 5$ control parameters. In such cases, any family of
functions is structurally stable and equivalent around
a critical point to one of these canonical forms.
Two symmetry-restricted catastrophe functions
(invariant under $x\to -x$), which are relevant to the present
contribution, are the $A_{+3}$ cusp catastrophe
with $x^4$ germ and the $A_{+5}$ butterfly catastrophe
with $x^6$ germ, given by
\bsub
\ba
&& A_{+3}\;\;\; {\rm cusp}\qquad\quad\;\;\, x^4 + a_2\,x^2 ~,
\label{cusp}
\\
&& A_{+5}\;\;\; {\rm butterfly}\qquad x^6 + a_4\,x^4 + a_2\,x^2 ~.
\label{but}
\ea
\label{cat-but}
\esub
In what follows, we study classical and quantum aspects of these
catastrophes in the algebraic framework of the interacting boson
model (IBM).

\section{Cusp catastrophe in the IBM}

A comprehensive classical analysis of shapes and stability, based
on the catastrophe theory formalism, has been carried out
in~\cite{castanos96} for the general IBM Hamiltonian with
one- and two-body terms. In the present section, we focus
the discussion to such interaction terms that, in addition,
conserve the SO(5) symmetry.
In this case, the energy surface of
Eq.~(\ref{enesurf}) is independent of $\gamma$  and has the
form $E_N(\beta,\gamma) = N(N-1)[z_0 + E(\beta)]$, where
\ba
E(\beta) =  (1+\beta^2)^{-2}[\, c\beta^4 + a\beta^2\,]\;\;\;,
\;\;\; c>0 ~.
\label{Ebeta12}
\ea
The coefficient $c$ is positive
to ensure stability at $\beta\to\infty$, and can be
chosen as a scale. The surface then
depends on one essential parameter $a$. The extremal points
occur at $\beta=0$ which is a minimum (maximum) for $a>0$
($a<0$), and at $\bs^2\!=\!-\tfrac{a}{(2c-a)}$ which
is a minimum (maximum) for $a<0$ ($a>2c>0$).

In the terminology of CT, $E(\beta)$ of Eq.~(\ref{Ebeta12})
corresponds to a cusp catastrophe with $\beta^4$ germ.
There is no Maxwell set and the bifurcation set consists of
a single~point
\ba
{\rm Critical\, point} \qquad\quad a=0~,
\label{cp}
\ea
which defines the critical point.
The corresponding critical surface is quartic,
\ba
E_{\rm cri}(\beta) =  c(1+\beta^2)^{-2}\beta^4 ~.
\label{Ecri-b4}
\ea
The phase diagram, shown in Fig.~5, is one-dimensional in the
control parameter space and involves spherical and deformed
regions, separated by the critical point.
Representative energy curves in each region
portray a single spherical minimum for $a>0$
which becomes flat at the critical point ($a=0$) and
evolves continuously into a deformed minimum
for $a<0$. These are characteristic features of a second order
phase transition.
\begin{figure}[t]
\centering
\begin{overpic}[width=8.5cm]{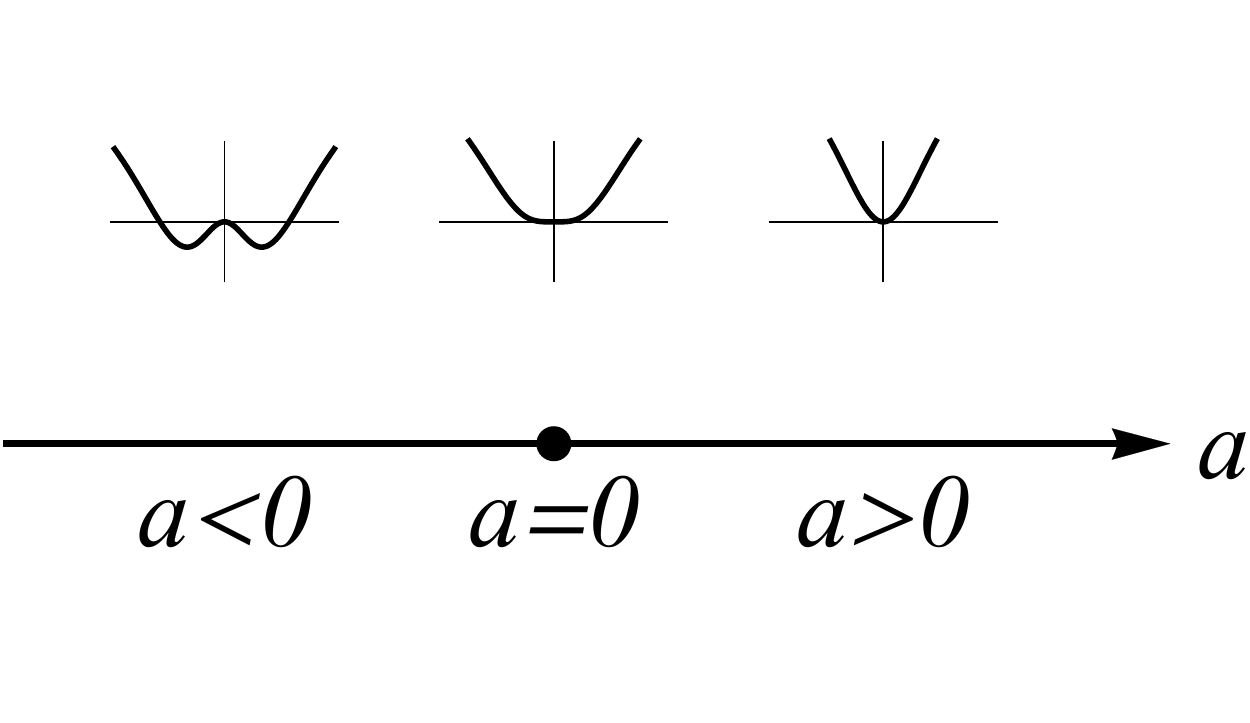}
\put (8,26) {\bf deformed}
\put (32,26) {\bf critical-point}
\put (64,26) {\bf spherical}
\end{overpic}
\vspace{-1cm}
\caption{\label{fig5-phase-diag-cusp}
\small
Phase diagram for the cusp catastrophe~(\ref{Ebeta12}) and representative
energy curves in each region.}
\end{figure}

The quantum properties of the cusp catastrophe can be studied
in the IBM by considering Hamiltonians whose energy surface
coincides with that of Eq.~(\ref{Ebeta12}). They are composed
of interaction terms associated with the U(5) and SO(6) DS
chains, Eqs.~(\ref{U5-ds})-(\ref{SO6-ds}),
\ba
{\rm U(6)}\supset
\left\{\begin{array}{c}
{\rm U(5)}\\
{\rm SO(6)}
\end{array}
\right\} \supset SO(5)\supset SO(3) ~.
\label{u5-o6}
\ea
The U(5)-SO(6) QPT has been studied extensively in the IBM
framework~\cite{iaczam04,Arias03,levgin03,Rowe04,dusuel05,
ramos05,cejnar06a,cejnar06b}, using an Hamiltonian of the form
\ba
\hat{H} = \epsilon\,\hat{n}_d + A\,P^{\dag}_0 P_0 ~.
\label{Hu5o6}
\ea
Here 
$\hat{n}_d \!=\! \hat{C}_{1}[{\rm U(5)}]$ is the linear Casimir operator
of U(5) with eigenvalues $n_d$, and
$P^{\dag}_0 P_0 \!=\! -\hat{C}_2[{\rm SO(6)}] + \hat{N}(\hat{N}+4)$
is related to the quadratic Casimir of SO(6)
with eigenvalues $\sigma(\sigma+4)$.
The operator $P^{\dag}_0 \!=\! d^{\dag}\cdot d^{\dag} - (s^{\dag})^2$ is
invariant under SO(6) and the centered dot denotes
a scalar product.
The fact that the two chains in Eq.~(\ref{u5-o6}) have a common 
$SO(5)\supset SO(3)$ segment,
implies that the eigenstates of $\hat{H}$ have good quantum
numbers $(N,\tau,n_{\Delta},L)$ for any choice of parameters.
For $\epsilon\!=\!0$ they coincide with
the U(5) basis states, $\ket{N,n_d,\tau,n_{\Delta},L}$~(\ref{U5-ds}),
for $A\!=\!0$ they are the SO(6) basis states,
$\ket{N,\sigma,\tau,n_{\Delta},L}$~(\ref{SO6-ds}) and, in general, their
wave functions can be expanded in both bases.

The above Hamiltonian has the energy surface of
Eq.~(\ref{Ebeta12}), with coefficients
$z_0\!=\!A$, $a \!=\! \tilde{\epsilon} - 4A$
and $c \!=\! \tilde{\epsilon}$, where
$\tilde{\epsilon} \!=\! \epsilon/(N-1)$.
The control parameter in $\hat{H}$
can be taken to be $\xi=A/\tilde{\epsilon}$, for which
$a=\tilde{\epsilon}(1-4\xi)$ and $c=\tilde{\epsilon}$.
The energy surface $E(\xi;\beta)$ and the Hamiltonian
$\hat{H}(\xi)$ depend parametrically on $\xi$.
The choice $a=0$, Eq.~(\ref{cp}), determines the critical-point
($\xi=\xi_c=1/4$), the critical surface
$E_{\rm cri}=E(\xi_c;\beta)$, Eq.~(\ref{Ecri-b4}), and the
critical Hamiltonian $\hat{H}_{\rm cri}=\hat{H}(\xi_c)$.

The spectrum of $\hat{H}_{cri}$ for the lowest $L=0,2,4,6$,
states and $E2$ transition rates between them, are shown
in Fig.~6(a). They exhibit features similar to those of the
E(5) critical-point symmetry~\cite{E5}, shown in Fig.~6(b).
The latter is obtained by an analytic solution of the Bohr
Hamiltonian, Eq.~(\ref{Bohr}), with a $\gamma$-independent
infinite square well potential $V(\beta)$. This similarity
is expected, since the surface at the critical point,
Eq.~(\ref{Ecri-b4}), has a flat behavior hence
can be replaced, to a good approximation, by such $V(\beta)$
for the low-lying states. A comparison of Fig.~6 with Fig.~2,
reveals that these spectra are intermediate between the
U(5)-DS and SO(6)-DS limits, reflecting the dynamics of
a flat-bottomed potential in-between the potentials of Fig.~1(a)
and Fig.~1(b) with a pronounced spherical and deformed minimum,
respectively. Experimental examples of such an E(5)-like structure
have been observed in nuclei~\cite{CasZam00}.

It is instructive to examine the structure of the wave
functions at and away from the critical point. Fig.~7 shows
the U(5) $n_d$-decomposition (left panels) and the SO(6)
$\sigma$-decomposition (right panels)
for eigenstates of $\hat{H}_{\rm cri}$, in comparison with the
structure at the U(5)-DS and SO(6)-DS limits.
For the U(5)-DS (bottom row panels), $n_d$ is a good
quantum number and each eigenstate exhibits a single
$n_d$ component, as expected for states of a spherical-vibrator.
The same eigenstates contain a mixture of different
$\sigma$-components, when expanded in the SO(6) basis, which
illustrates that U(5) and SO(6) are incompatible symmetries.
For the SO(6)-DS (upper row panels), the eigenstates are pure
with respect to the SO(6) quantum number $\sigma$, and their
U(5) decomposition exhibits a broad $n_d$-distribution,
as expected for states of a deformed $\gamma$-unstable rotor.
The eigenstates of $\hat{H}_{\rm cri}$ (middle row panels),
have a structure in-between these DS limits.
On one hand, they contain several U(5) $n_d$-components but,
on the other hand,  they are not as fragmented as the
SO(6) basis states. Their SO(6) $\sigma$-distribution is
broad but is peaked at a lower value of $\sigma$, compared
to the U(5) basis states.
\begin{figure}[t]
\centering
\hspace{-0.56cm}
\begin{overpic}[width=8.3cm]{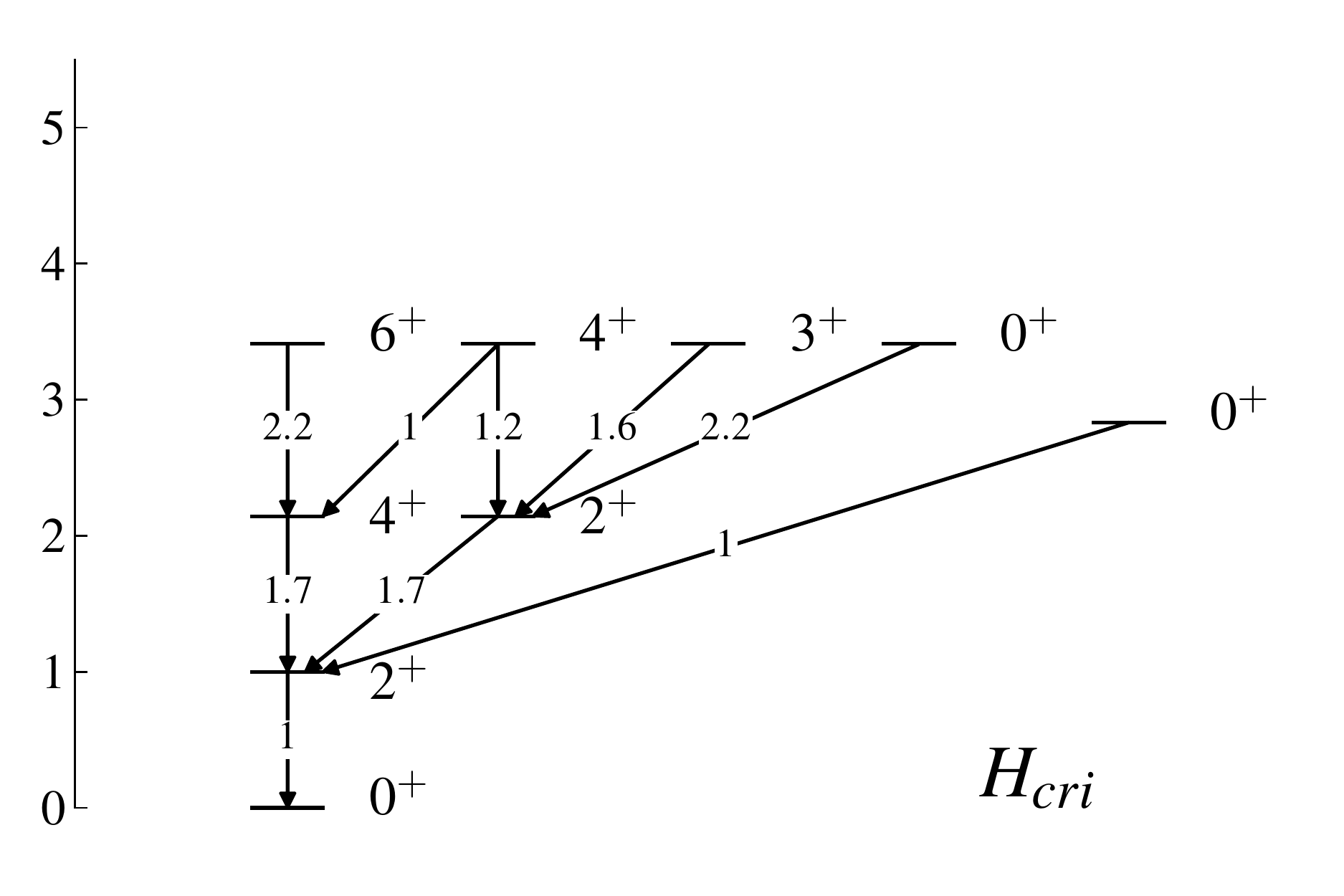}
\put (63,7.8) {\large(a)}
\end{overpic}
\hspace{-0.4cm}
\begin{overpic}[width=8.3cm]{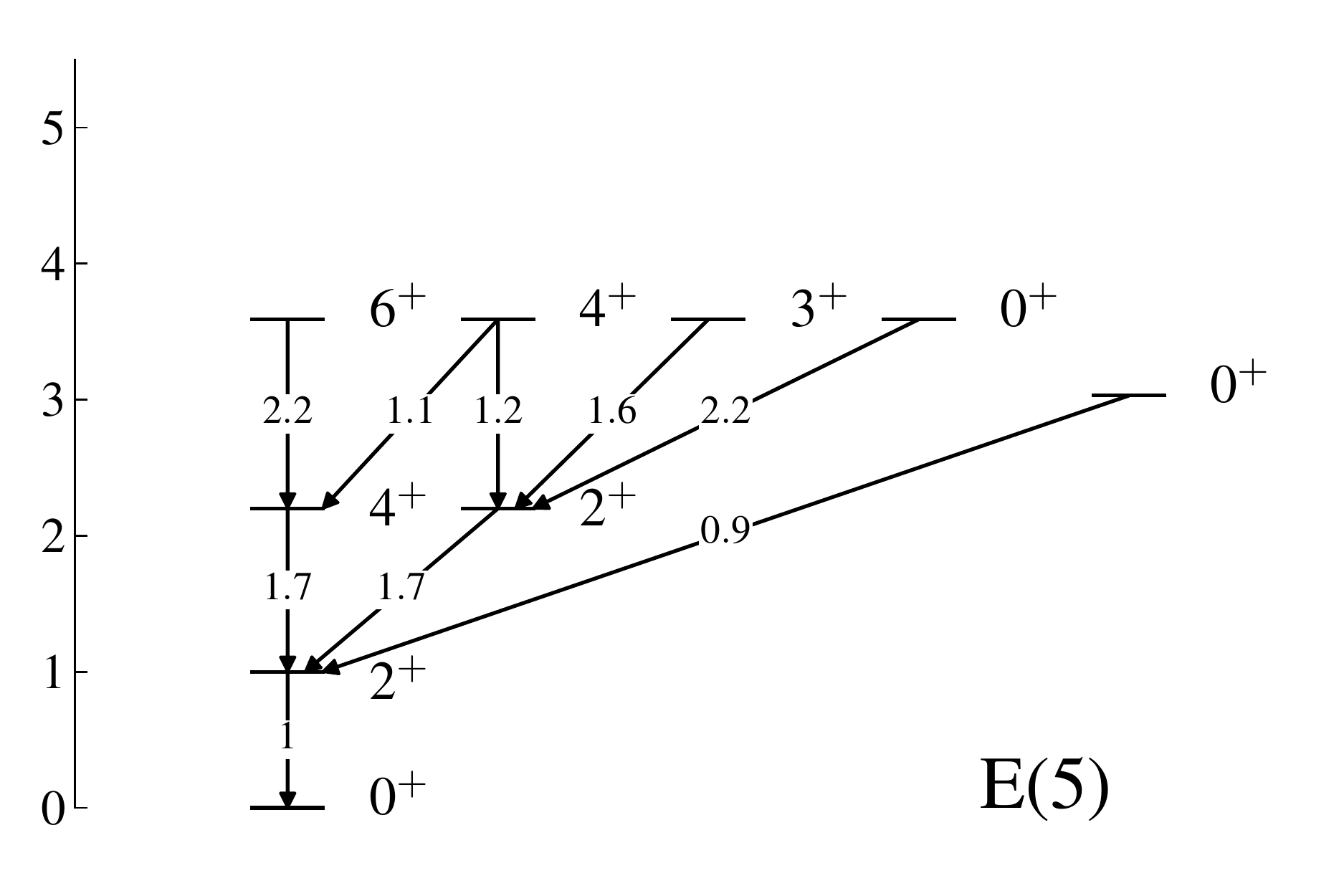}
\put (63,6.5) {\large(b)}
\end{overpic}
\caption{\label{fig6-E5IBM-E5}
\small
Energy spectra and $E2$ rates in units as in Fig.~2.
(a)~The algebraic critical-point Hamiltonian $\hat{H}_{\rm cri}$,
obtained from Eq.~(\ref{Hu5o6}) with $\epsilon = 4A(N-1)$ and
$N=25$. The $E2$ operator is proportional to
$d^{\dag}_{m}s +  s^{\dag}\tilde{d}_{m}$.
(b)~The geometric E(5) critical-point Hamiltonian~\cite{E5},
obtained from Eq.~(\ref{Bohr}) with an infinite square-well
potential $V(\beta)$. The $E2$ operator is proportional to
$\alpha_{2,m}$, Eq.~(\ref{radius}).}
\label{fig2}
\end{figure}
\begin{figure}[t]
\centering
\includegraphics[width=16cm]{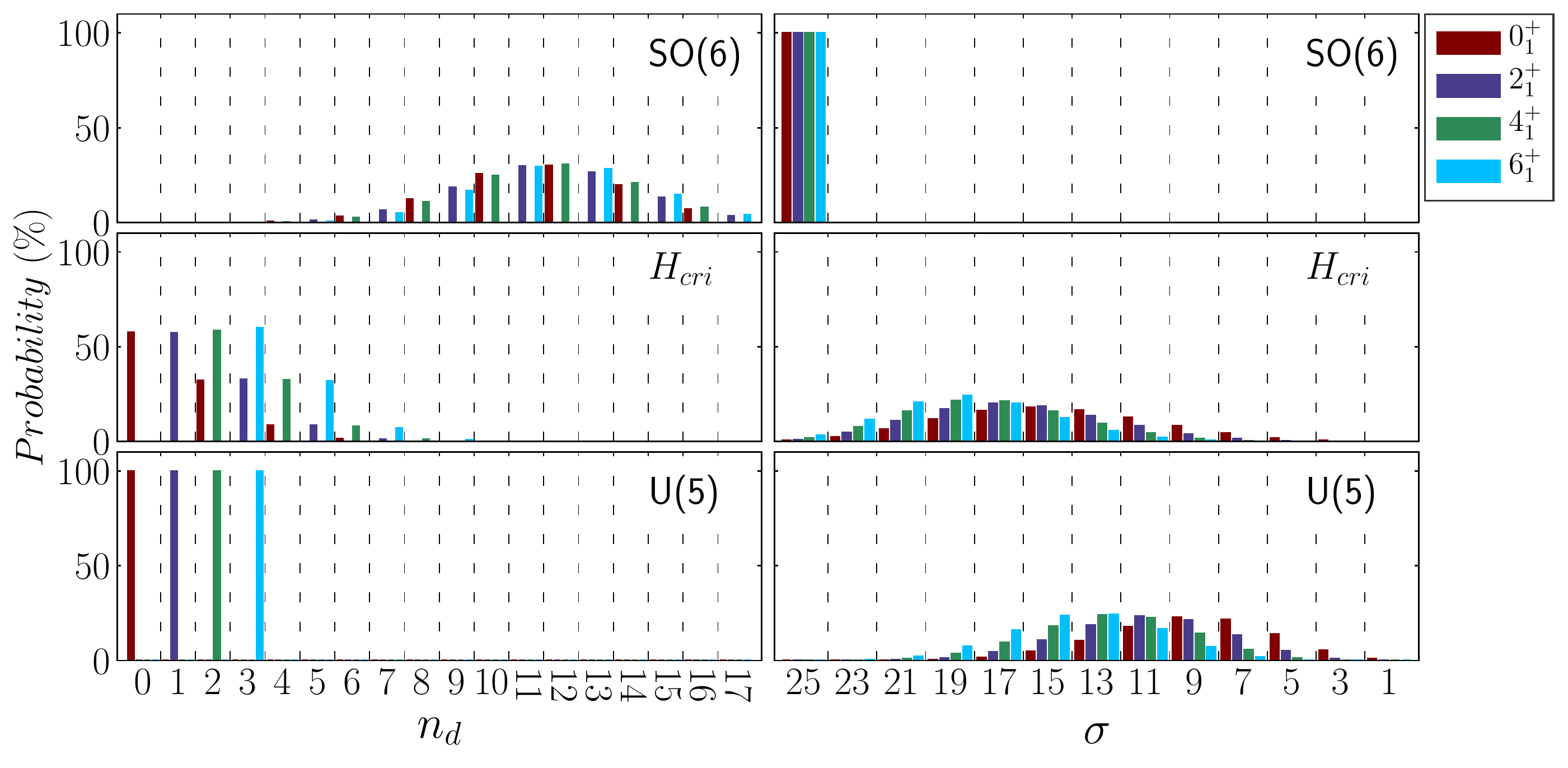}
\caption{\label{fig7-decomp-u5-o6-cusp}
\small
U(5) $n_d$-decomposition and SO(6) $\sigma$-decomposition for
selected eigenstates of the U(5)-DS Hamiltonian
[Fig.~2(a)], the SO(6)-DS Hamiltonian [Fig.~2(b)] and the
critical Hamiltonian~[Fig.~6(a)].}
\end{figure}

\section{Butterfly catastrophe in the IBM}

The cusp catastrophe of Eq.~(\ref{Ebeta12}) is obtained with
SO(5)-invariant one- and two-body terms in the IBM Hamiltonian,
and is suitable for describing second order QPTs between
spherical and $\gamma$-unstable deformed shapes.
In order to accommodate first-order transitions involving a
coexistence of such shapes, one needs to have higher powers
of $\beta^2$ in the Landau potential. This can be accomplished
by including higher-order terms in the IBM Hamiltonian.
In the present section, we consider the effects of including
such cubic terms which conserve the SO(5)
symmetry~\cite{gavlev20}.
In this case, the energy surface of Eq.~(\ref{enesurf})
has the form $E_N(\beta,\gamma) = N(N-1)(N-2)[z_0 + E(\beta)]$,
where
\ba
E(\beta) =  (1+\beta^2)^{-3}[\,A\beta^6 + D\beta^4 + F\beta^2\,]
\;\;\;,\;\;\; A>0 ~.
\label{Ebeta123}
\ea
The positive coefficient $A$ can be taken as a scale, hence the 
surface depends on two essential parameters $D$ and $F$.
The extremal points of the surface occur at $\beta=0$ and at
\bsub
\ba
\rho_{\pm} &=& \beta^{2}_{\pm} =
\frac{F-D\pm\sqrt{\Delta}}{3A-D} ~,
\label{rhopm}\\
\Delta &=& (F-D)^2-(3A-D)F ~.
\label{Delta}
\ea
\label{defsol}
\esub
The spherical $\beta=0$ extremum 
is a local minimum (maximum) for $F>0$ ($F<0$) and becomes
a global minimum for $F>0$ and $D^2-4AF<0$.
$\rho_{\pm}$ of Eq.~(\ref{rhopm}) are solutions of a quadratic
equation in the variable $\rho=\beta^2$, hence are physically
acceptable if positive.
When such deformed extrema exist, then $\rho_{+}>0$
($\rho_{-}>0$) is a local minimum (maximum) and $\rho_{+}>0$
becomes a global minimum for $D^2-4AF>0$.

In the terminology of CT, $E(\beta)$ of Eq.~(\ref{Ebeta123})
corresponds to a butterfly catastrophe with $\beta^6$ germ.
The phase diagram, shown in Fig.~8,
is two-dimensional in the control parameter space $(D,F)$,
and involves
(I)~spherical, (II)~deformed and (III)~coexistence regions.
Region~(I) defines the spherical phase, in which the
energy surface has as a single minimum at $\beta=0$.
It is composed of three sub-regions
I(a)~${\textstyle 3A-D<0,F>0}$, [$\rho_{-}\!>\!0$ (max),
$\rho_{+}\!<\!0$];
I(b)~${\textstyle 3A-D>0,D\!>F>0,\Delta>0}$,
[$\rho_{-}\!<\!\rho_{+}\!<\!0$];
I(c) $\Delta\!<\!0$, [$\rho_{-},\rho_{+}$ complex].
Region (II) defines the deformed phase, in which the
energy surface has a single deformed minimum at
$\rho_{+}\!>\!0$.
It is composed of two sub-regions:
II(a)~${\textstyle 3A-D<0,F<0,\Delta>0}$,
[$\rho_{-}\!<\!0$, $\rho_{+}\!>\!0$ (min)];
II(b)~${\textstyle 3A-D>0,F<0}$, [$\rho_{-}\!>\!0$ (max),
  $\rho_{+}\!>\!0$ (min)];
Region (III) defines the coexistence phase, in which the
energy surface has both spherical and deformed minima
at $\beta\!=\!0$ and $\rho_{+}\!>\!0$, respectively.
It is composed of two sub-regions
III(a)~${\textstyle 3A-D>0,F>0,\Delta>0,D^2-4AF<0}$,
and III(b)~$D<0,F>0,\Delta>0, D^2-4AF>0$,
[$\rho_{-}>0$ (max), $\rho_{+}>0$ (min)].
\begin{figure}[t]
\centering
\includegraphics[width=10cm]{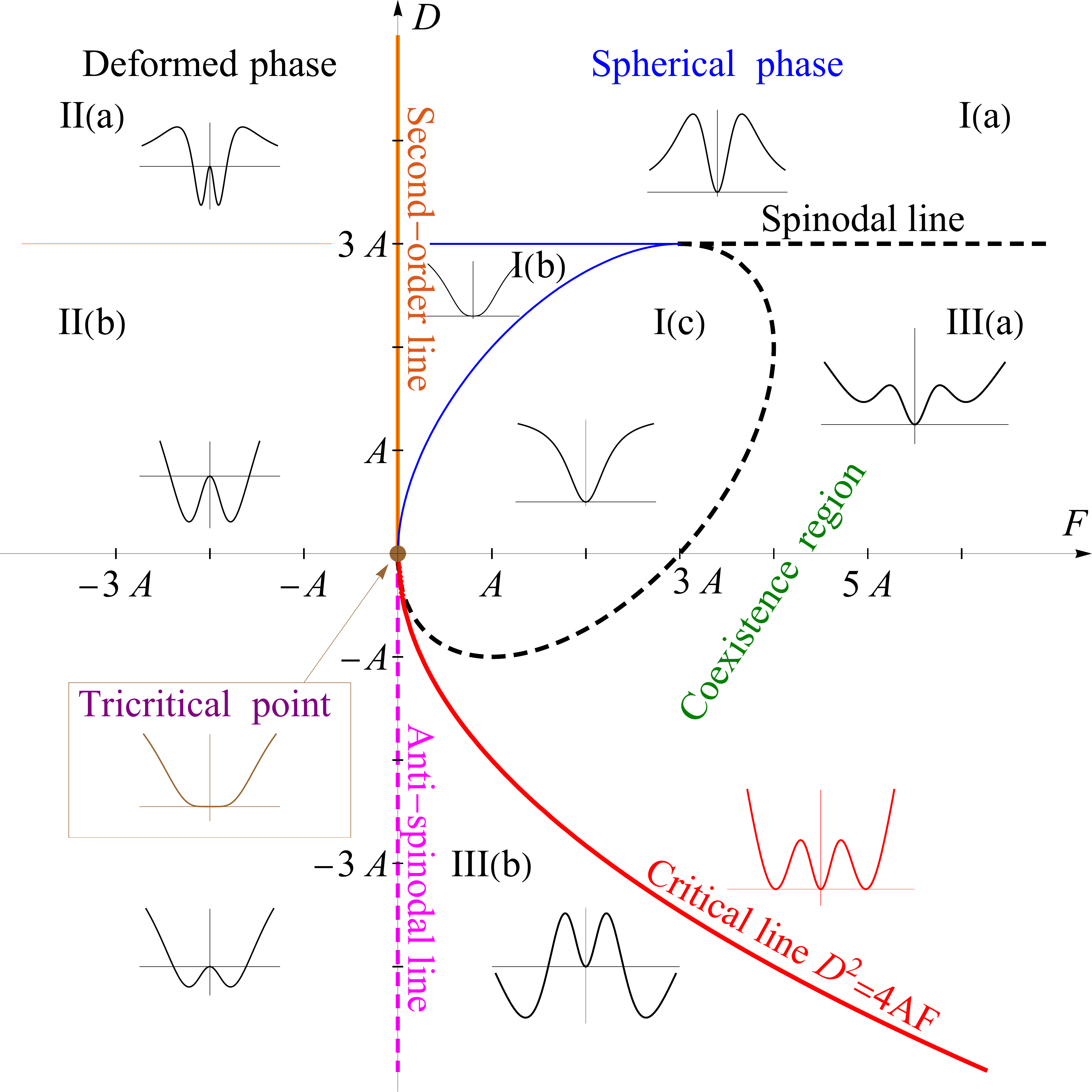}
\caption{\label{fig8-phase-diag-butt}
\small
Phase diagram for the butterfly catastrophe~(\ref{Ebeta123}),
composed of (I)~spherical, (II)~deformed, and (III)~coexistence
regions, and representative energy curves within each region.
The bifurcation set, Eq.~(\ref{bifur}), and Maxwell
set, Eq.~(\ref{maxwell}), separating the different regions,
are depicted.}
\end{figure}

The bifurcation sets are given by
\bsub
\ba
&&{\rm Spinodal\, line}\qquad \qquad\; \Delta=0,\; F>D\; 
; \; D=3A,\; F>D ~,
\label{spinod}\\
&&{\rm Anti\!-\!spinodal\, line}\qquad F=0,\; D<0 ~,
\label{antispinod}\\
&&{\rm Second\, order\, line}\qquad\quad  F=0,\; D>0 ~,
\label{2nd}\\
&&{\rm Tricritical\, point}\qquad\quad\;\, F=D=0 ~.
\label{tricrit}
\ea
\label{bifur}
\esub   
The Maxwell set defines the critical line
\ba
&&{\rm Critical\, line} \qquad\quad D^2-4AF=0,\quad F>0,\,D<0 ~,
\label{maxwell}
\ea
along which the critical energy surface acquires the form
\ba
E_{\rm cri}(\beta) =  A(1+\beta^2)^{-3}\beta^2(\beta^2-\b0^2)^2 ~,
\label{Ecri-b6}
\ea
where $\b0^2 \!=\! -D/2A\!=\! -2F/D$.
$E_{\rm cri}(\beta)$ has two degenerate minima,
at $\beta=0$ and $\beta_{+}^2=\b0^2$, and a maximum at
$\beta_{-}^2=\tfrac{\b0^2}{3+2\b0^2}$. The two minima
are separated by a barrier of height
$h=A\tfrac{4\b0^6}{27(1+\b0^2)}$.
The surface landscape is similar to that
of Fig.~1(c), and its profile is shown in Fig.~9.  
\begin{figure}[b]
\centering
\includegraphics[width=6cm]{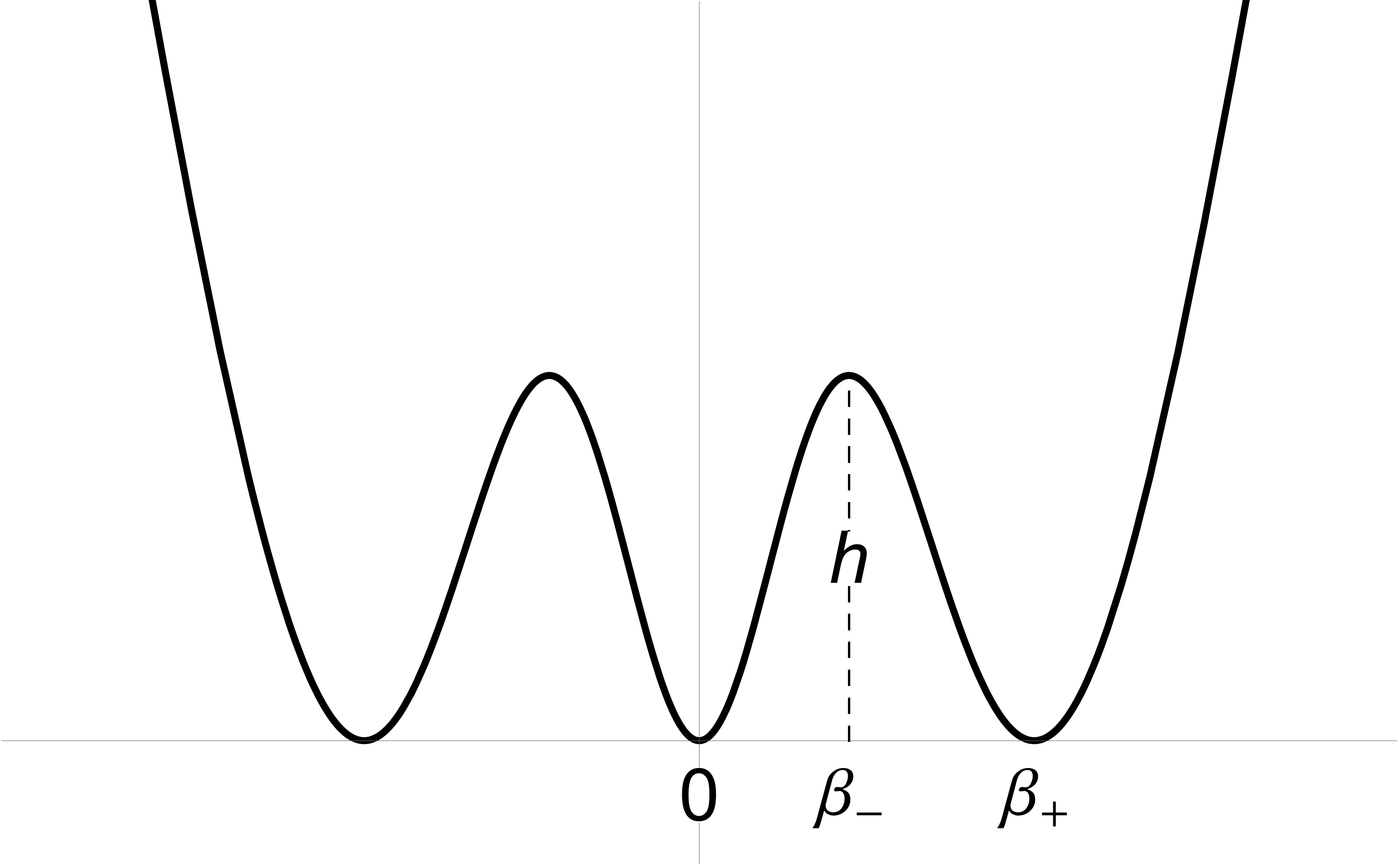}
\caption{\label{fig9-barrier}
\small
Profile of the critical energy surface
$E_{\rm cri}(\beta)$, Eq.~(\ref{Ecri-b6}).
$\beta\!=\!0$ and $\beta_{+}$ are two degenerate minima,
$E_{\rm cri}(0)\!=\!E_{\rm cri}(\beta_{+})\!=\!0$.
The position and height of the barrier are
$\beta_{-}$ and $h\!=\!E_{\rm cri}(\beta_{-})$.}
\end{figure}

The bifurcation and Maxwell sets comprise the separatrix
and determine the boundaries of the different regions
in the phase diagram of Fig.~8.
The spinodal line, Eq.~(\ref{spinod}),
separates the spherical and coexistence regions.
The anti-spinodal line, Eq.~(\ref{antispinod}), separates
the coexistence and deformed regions.
The second-order line, Eq.~(\ref{2nd}),
separates the deformed and spherical regions.
The critical line, Eq.~(\ref{maxwell}),
partitions the coexistence region
to a sub-region III(a), where the spherical (deformed)
minimum is global (local), and a sub-region~III(b), where
the two minima exchange roles.
All lines intersect at the tricritical point,
Eq.~(\ref{tricrit}),
for which $E(\beta) =  A(1+\beta^2)^{-3}\beta^6$.

As the control parameters are varied and the separatrix
lines are crossed, the energy surface typography exhibits
a qualitative change and a shape-phase transition occurs.
From Fig.~8 one sees that across the second-order
line, as $F$ increases,
a single deformed minimum ($F\!<\!0$) changes
continuously into a single spherical minimum ($F\!>\!0$),
in accord with a second-order transition.
Across the spinodal line, as $D$ decreases,
a second local deformed minimum occurs. The two minima
become degenerate at the critical line and then cross.
The deformed minimum becomes global
and the coexistence of the two minima persists
until the local spherical minimum disappears
at the anti-spinodal line. These are the attributes of
a first-order transition. Interestingly, across the
anti-spinodal line, as $F$ decreases from positive to
negative values, the deformed
minimum remains global but $\beta=0$ changes from
a local minimum to a maximum, a~feature encountered
in a second-order transition.

The quantum properties of the butterfly catastrophe can be
studied in the IBM, by considering Hamiltonians whose
energy surface coincides with that of Eq.~(\ref{Ebeta123}).
Since the latter is sextic, this requires higher-order terms
in the Hamiltonian.
For that purpose, we extend the one- and two-body
Hamiltonian of Eq.~(\ref{Hu5o6}), by adding to it a cubic term
composed of invariant operators of U(5) and SO(6).
The resulting Hamiltonian has the form~\cite{gavlev20},
\ba
\hat{H} = \epsilon\,\hat{n}_d + r\,P^{\dag}_0 P_0
+ h_2\,P^{\dag}_0\hat{n}_d P_0 ~.
\label{Hu5o6-cubic}
\ea
Since both SO(5) and SO(3) are preserved by the above
Hamiltonian, its eigenstates have good $(\tau,L)$ quantum
numbers and can be labeled as $L^{+}_{i,\tau}$, where the
ordinal number $i$ enumerates the occurrences of states
with the same $(\tau,L)$, with increasing energy.

One of the three parameters in $\hat{H}$~(\ref{Hu5o6-cubic})
can be taken as a global scale and we set $h_2\!=\!1$.
The resulting energy surface, Eq.~(\ref{Ebeta123}),
has coefficients: $z_0\!=\!\bar{r}$, $A\!=\!1+\bar{\epsilon}$,
$D\!=\! -2(1+2\bar{r}-\bar{\epsilon})$ and
$F\!=\! 1-4\bar{r} +\bar{\epsilon}$,
where $\bar{\epsilon}\!=\!\epsilon/(N-1)(N-2)$ and
$\bar{r}\!=\!r/(N-2)$ are the essential control parameters.
The bifurcation and Maxwell sets of
Eqs.~(\ref{bifur}) and (\ref{maxwell}),
translate into the following conditions on $\bar{\epsilon}$
and $\bar{r}$ (assuming $\bar{r}\geq 0$), 
\bsub
\ba
&&{\rm Spinodal\, line}\qquad \qquad\;
\bar{\epsilon} = \tfrac{1}{3}(1+2\bar{r})^2 \;\;, \;\; \bar{r} < 1 ~,
\label{spinod2}\\
&&{\rm Anti\!-\!spinodal\, line}\qquad 
\bar{\epsilon} = 4\bar{r}-1 \;\;, \;\; \bar{r} < 1 ~,
\label{antispinod2}\\
&&{\rm Second\, order\, line}\qquad\quad
\bar{\epsilon} = 4\bar{r}-1 \;\;, \;\; \bar{r} > 1 ~,
\label{2nd2}\\
&&{\rm Tricritical\, point}\qquad\quad\;\,
\bar{\epsilon}=3 \;\;,\;\; \bar{r} =1 ~,
\label{tricric2}\\
&&{\rm Critical\, line} \qquad\qquad\quad
\bar{\epsilon} = \bar{r}(\bar{r}+2) \;\;, \;\; \bar{r} < 1 ~.
\label{cri-line}
\ea
\label{sepratrix}
\esub   
In particular, the parameters of the critical Hamiltonian,
$\hat{H}_{\rm cri}$, satisfy the condition of Eq.~(\ref{cri-line}).
The corresponding critical surface, $E_{\rm cri}(\beta)$, 
Eq.~(\ref{Ecri-b6}) and Fig.~9, 
has a maximum at $\beta_{-}^2$ and two degenerate minima
at $\beta=0$ and $\beta_{+}^2$ separated by a barrier of
height $h$, given by
\ba
\beta_{-}^2=\frac{1-\bar{r}}{5+\bar{r}} \;\;\;, \;\;\;
\beta_{+}^2=\frac{1-\bar{r}}{1+\bar{r}} \;\;\;, \;\;\;
h = \frac{2}{27}(1-\bar{r})^3 ~.
\label{barrier}
\ea
Small values of $\bar{r}$ ($\bar{r}\to 0$) correspond to
a high barrier, while large values ($\bar{r}\to 1$) correspond
to a low barrier.

The calculated spectrum of the critical Hamiltonian,
$H_{\rm cri}$, for $N=25$, is shown in Fig.~10.
The U(5) ($n_d$) and SO(6) ($\sigma$) decompositions of
selected eigenfunctions are portrayed in~Fig.~11.
For a low-barrier ($\bar{r}=0.9$), the pattern of energy
levels and $E2$ transitions, displayed in Fig.~10(a),
resembles an E(5)-like structure (compare with Fig.~6).
This interpretation is corroborated for the yrast states,
$L^{+}_{1,\tau}$ with $L\!=\!2\tau\!=\!0,2,4,6$, by their
$n_d$-decomposition shown in Fig.~11(d), and by
their $\sigma$-decomposition shown in Fig.~11(b).
Both decompositions are seen to be similar to those of
the critical ``cusp'' Hamiltonian (compare with the middle
panels of Fig.~7).
The $n_d$-decomposition for the non-yrast states,
$L^{+}_{2,\tau}$ with $L\!=\!2\tau=0,2,4$, shown in
Fig.~11(f), displays similar trends.
\begin{figure}[t]
\centering
\begin{overpic}[width=8cm]{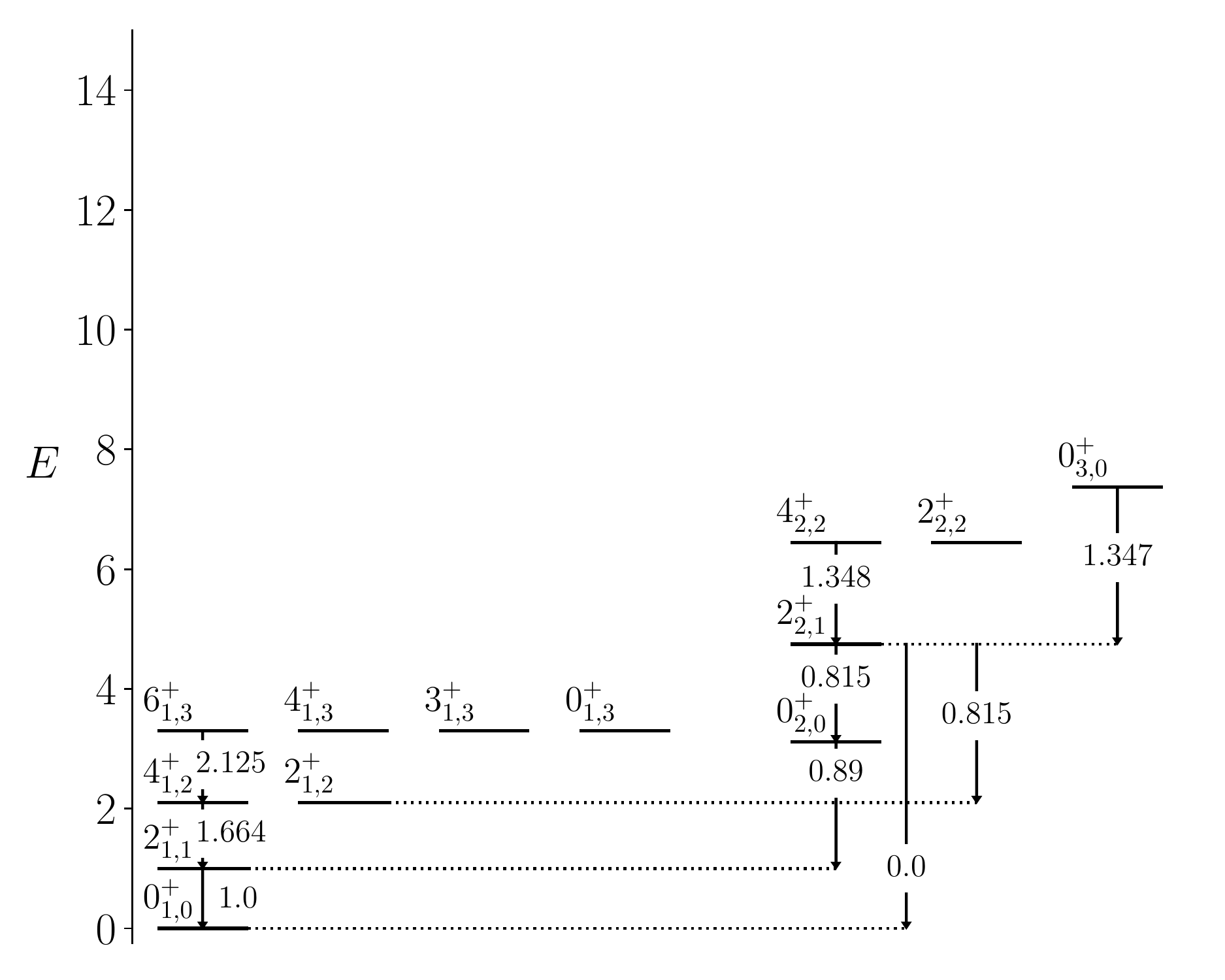}
\put (15,67) {(a) Low barrier}
\end{overpic}  
\hspace{-0.3cm}
\begin{overpic}[width=8cm]{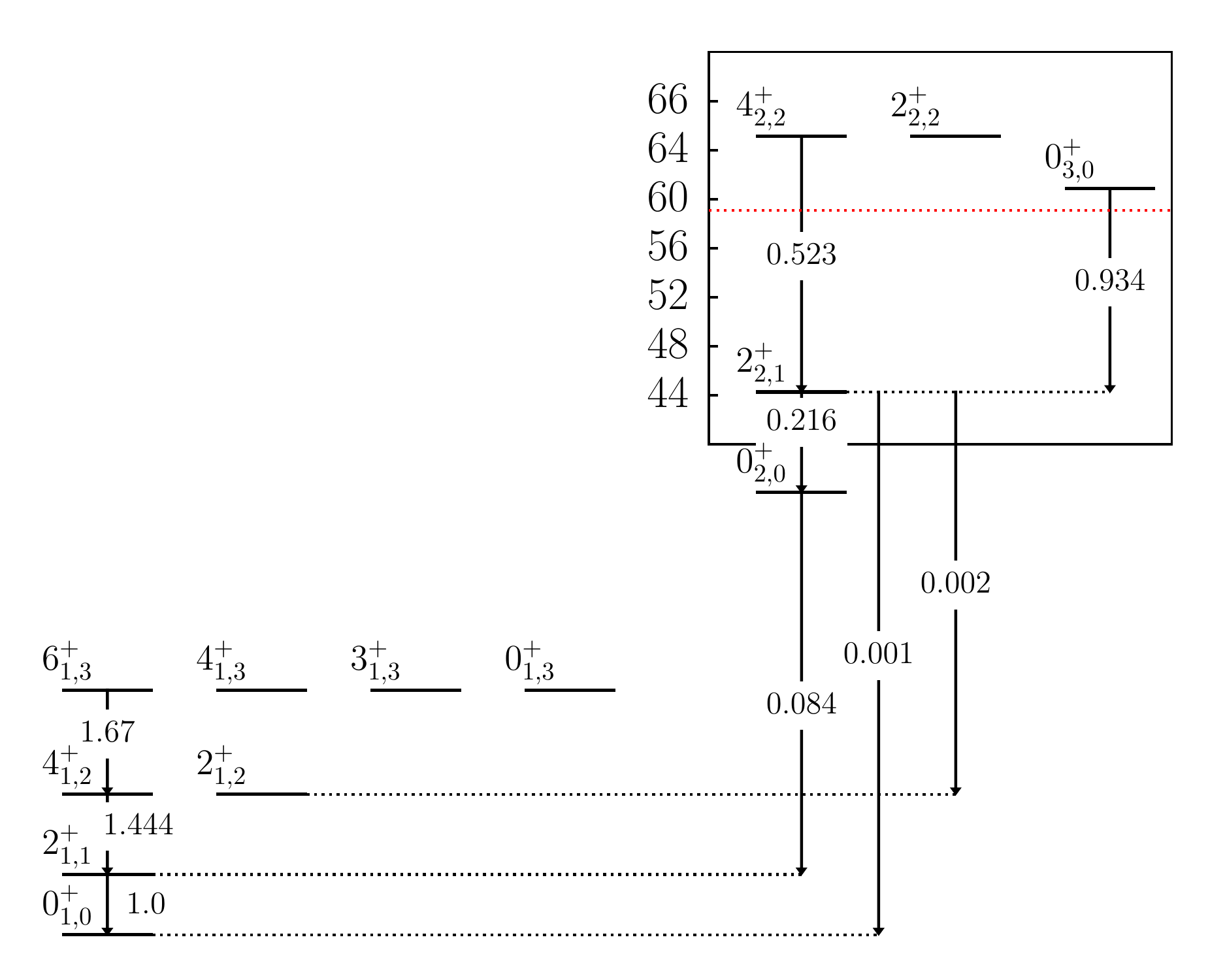}
\put (4,67) {(b) High barrier}
\end{overpic}  
\caption{\label{fig10-spec-low-high-bar}
\small
Energy spectra [in units of $E(2^{+}_{1,1})=1$] and $E2$ rates
[in units of $B(E2;2^{+}_{1,1}\to 0^{+}_{1,0})=1$]
for the critical Hamiltonian,
Eq.~(\ref{Hu5o6-cubic}) with parameters
as in Eq.~(\ref{cri-line}) and $N=25$.
The $E2$ operator is proportional to
$d^{\dag}s + s^{\dag}\tilde{d}$.
(a)~Low barrier (negligible height), $\bar{r}=0.9$.
(b)~High barrier, $\bar{r}=0.1$. The barrier's height
is indicated by a horizontal dotted line in the inset.}
\label{fig-spec-high-low-b}
\end{figure}
\begin{figure}[t]
\centering
\includegraphics[width=16cm]{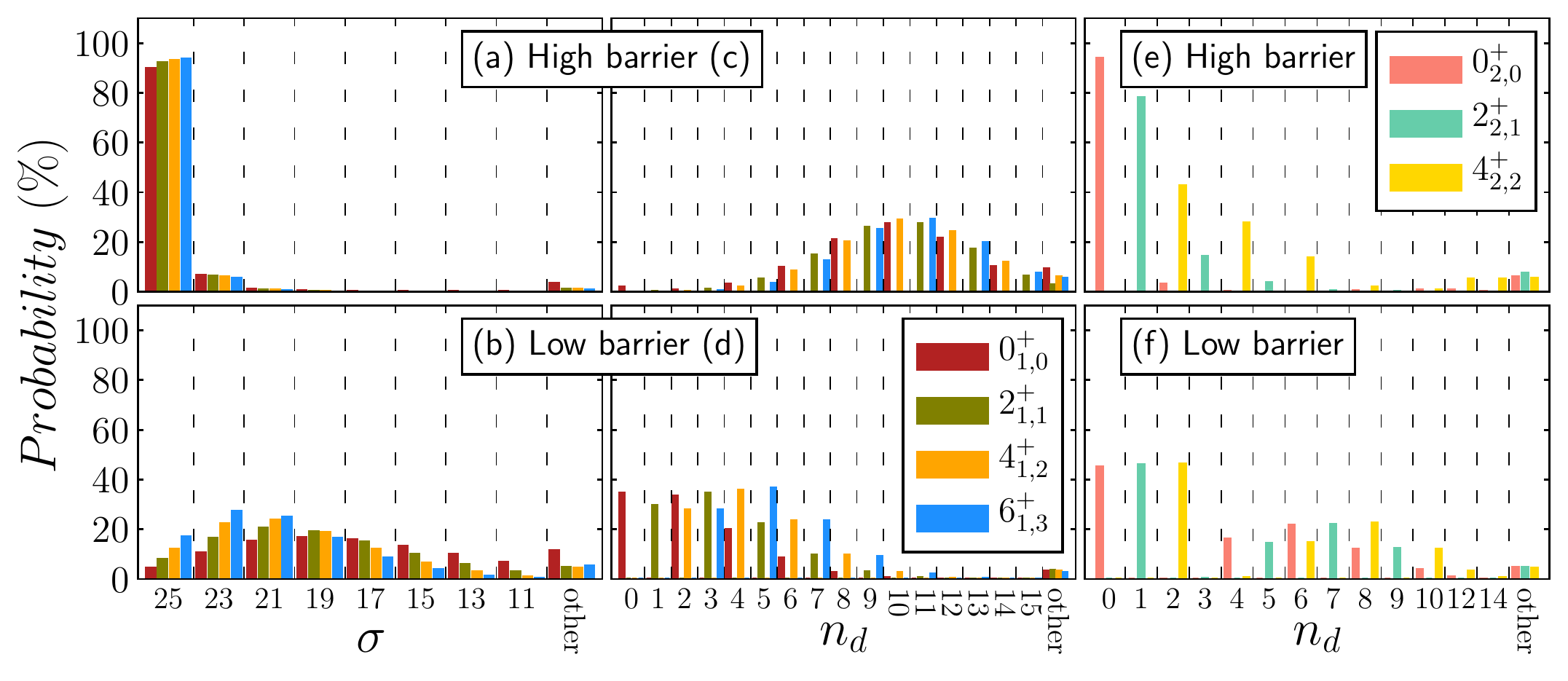}
\caption{\label{fig11-decomp-u5-o6-butt}
\small
U(5) ($n_d$) and SO(6) ($\sigma$) decompositions
for selected eigenstates of $\hat{H}_{\rm cri}$ [Fig.~10(a)],
appropriate for a low-barrier (bottom row) 
and [Fig.~10(b)], appropriate for a high-barrier
(top row). Panels (a)-(b)-(c)-(d) address
yrast states ($L^{+}_{1,\tau}$) and panels (e)-(f) address
non-yrast states ($L^{+}_{2,\tau}$).}
\end{figure}

For a high-barrier ($\bar{r}=0.1$), the spectrum of
$\hat{H}_{\rm cri}$ exhibits two distinct classes of states.
The first class consists of yrast states, $L^{+}_{1,\tau}$,
which form the ground band of a $\gamma$-unstable deformed rotor. 
As shown in Fig.~10(b), the pattern of energies and $E2$
transitions of these states is similar to that of the
SO(6)-DS limit (compare with Fig.~2).
The $\sigma$-decomposition shown Fig.~11(a), discloses that
these states posses the SO(6) quantum number
$\sigma\!=\!N$, to a good approximation. The broad
$n_d$ distribution shown in Fig~11(c),
discloses their deformed nature. These attributes are similar
to those encountered in the SO(6)-DS limit (compare with the
upper panels of Fig.~7).

The second class consists of non-yrast states, $L^{+}_{2,\tau}$,
arranged in $n_d$-multiplets of a spherical vibrator.
As shown in the inset of Fig.~10(b), the pattern of energies
and $E2$ transitions for these states is similar to that of
the U(5)-DS limit (compare with Fig.~2).
The $n_d$-decomposition, shown Fig.~11(e), discloses that their
wave functions are dominated by a single $n_d$ component
(especially, the $0^{+}_{2,0}$ and $2^{+}_{2,1}$ states),
similarly to the spherical U(5) basis states (compare with
the lower panels of Fig.~7).
The fact that for a high barrier, the lowest spherical-type
of states show a high degree of purity with respect to U(5)
and the lowest deformed-type of states, members of the ground
band, show a high degree of purity with respect to SO(6),
while other states of the same Hamiltonian are mixed with
respect to both U(5) and SO(6), highlights the relevance of
partial dynamical symmetries~\cite{Leviatan11,Leviatan96}
to the spectrum of the critical Hamiltonian.
Such a novel symmetry property was encountered
in numerous studies of single
shapes~\cite{Leviatan11,Leviatan96,LevSin99,GarciaRamos09,
  levgav12,Leviatan13} and multiple
shapes~\cite{Leviatan07,LevDek16,levgav17,levgav18,
  LevGavRamIsa18} in nuclei, and of systems
with mixed regular and chaotic
dynamics~\cite{walev93,levwhe96,Macek14}.

The evolution of the dynamics along the critical line,
from an E(5)-like
structure for a low-barrier, to coexisting SO(6)-like and
U(5)-like structure
for a high barrier, can be investigated by varying the
control parameter $\bar{r}$ in the critical Hamiltonian.
The results of such a detailed study will be
reported elsewhere~\cite{gavlev20}.
A particularly sensitive measure of the
height of the barrier is the ratio of $E2$ rates
$B(E2; 0^{+}_{2,0}\to 2^{+}_{1,1})/B(E2; 2^{+}_{1,1}\to 0^{+}_{1,0})$.
As indicated in Fig.~10, this ratio is $0.89$ ($0.084$)
for $\bar{r}=0.9$ ($\bar{r}=0.1$), {\it i.e.}, an
order-of-magnitude change from low to high barrier.

\section{Concluding remarks}

We have employed the algebraic framework of the
interacting boson model (IBM) for the study of
quantum catastrophes. The latter are manifested
in first- and second-order QPTs between spherical and
$\gamma$-unstable deformed nuclear shapes,
whose dynamics is described by the U(5) and SO(6)
dynamical symmetries (DSs) of the IBM, respectively.

The classical analysis involves the construction of the
Landau potentials by means of coherent states, and the
study of the complete phase diagrams
by the methods of Catastrophe Theory. The potentials
depend on the quadrupole shape variable $\beta$, and on
control parameters of the Hamiltonian. A quartic-type of
potential exemplifying a cusp catastrophe and a sextic type
of potential exemplifying a butterfly catastrophe,
have been considered. The relevant bifurcation and
Maxwell sets have been identified. The latter mark the
boundaries of different regions in the phase diagram,
encompassing classes of potentials with similar typography.
The crossing of these boundaries corresponds to
a shape-phase transition.

The quantum analysis involves a study of the spectra
and the structure of wave functions,
arising from IBM Hamiltonians which mix terms from two
incompatible DS chains. For the cusp catastrophe,
one- and two-body interaction terms are sufficient and
the Hamiltonian is expanded in terms of the Casimir operators
of the leading sub-algebras of the two chains, 
$\hat{H}(a_1,a_2) = a_1\,\hat{C}_{k}[G_1]
+ a_2\,\hat{C}_{n}[G_2]$.
For the butterfly catastrophe, a three-body term is required
and the Hamiltonian involves a general expansion in terms of
invariant operators of ${\rm G_1}$ and ${\rm G_2}$,
$\hat{H}(a_{n,k})
= \sum_{n,k} a_{n,k}\,{\cal I}_n[G_1]\, {\cal I}_k[G_2]$.
In the examples considered,
${\rm G_1=U(5)}$ and ${\rm G_2=SO(6)}$.

The symmetry-restricted cusp catastrophe is appropriate
for a second order QPT. The phase diagram is
one-dimensional, involving spherical and deformed regions
corresponding to potentials with single minima, and
a critical point corresponding to a flat-bottomed
quartic potential.
The quantum dynamics exhibits a continuous evolution
from a U(5) structure to a SO(6) structure, with
an E(5)-like structure at the critical point.
The symmetry-restricted butterfly catastrophe can
accommodate both second- and first-order QPTs.
The phase diagram is two-dimensional involving
spherical and deformed regions, where the potential has
a single minimum, and a coexistence region,
where two minima occur in the potential.
The quantum dynamics along the
critical line, where the two minima are degenerate,
displays an evolution from an E(5)-like structure
appropriate to a low-barrier to a coexistence of
U(5)-like and SO(6)-like structure for a high barrier.
A symmetry analysis of the wave functions for the high
barrier case, reveals the coexistence of U(5) and
SO(6) partial dynamical symmetries.
\ack
This work is supported by the Israel Science Foundation.

\end{document}